\documentclass[10pt,journal,final,twoside,letterpaper]{%
  IEEEtran}

\pdfoutput=1

\usepackage{amsmath, amssymb, amsfonts, amsthm}
\usepackage{graphicx,epsfig,epstopdf}
\usepackage{multirow,rotating}
\usepackage[noadjust]{cite}
\usepackage{floatflt}
\usepackage{subfigure}
\usepackage{url}
\usepackage{calc,ifthen}
\usepackage{caption}
\usepackage{psfrag}
\usepackage{cases}
\usepackage{pifont}
\usepackage{boxedminipage}
\usepackage{balance, blkarray}
\usepackage{booktabs} 
\newtheorem{mydef}{Definition}
\newtheorem{mylem}{Lemma}
\newtheorem{mythm}{Theorem}
\newtheorem{myprop}{Proposition}
\newtheorem{mycoro}{Corollary}
\newcommand{\tb}{\textbf}
\newcommand{\mb}{\mathbf}
\newcommand{\mc}{\mathcal}
\newcommand{\ol}{\overline}
\def\QEDclosed{\mbox{\rule[0pt]{1.3ex}{1.3ex}}} 
\def\QEDopen{{\setlength{\fboxsep}{0pt}\setlength{\fboxrule}{0.2pt}\fbox{\rule[0pt]{0pt}{1.3ex}\rule[0pt]{1.3ex}{0pt}}}}

\begin{document}


\title{%
  Consensus in the Presence of Multiple Opinion Leaders: Effect of Bounded Confidence}

\author{%
  Ranga~Dabarera,~\IEEEmembership{Student~Member,~IEEE,} 
  Kamal~Premaratne,~\IEEEmembership{Senior~Member,~IEEE,} 
  Manohar~N.~Murthi,~\IEEEmembership{Member,~IEEE,} 
  and~Dilip~Sarkar,~\IEEEmembership{Senior~Member,~IEEE \\
  \hspace*{0in} \\
  \begin{minipage}[t]{7.0in}{\small{\textbf{\copyright 2016 IEEE. Personal use of this material is permitted. Permission from IEEE must be obtained for all other uses, in any current or future media, including reprinting/republishing this material for advertising or promotional purposes, creating new collective works, for resale or redistribution to servers or lists, or reuse of any copyrighted component of this work in other works.}}}
  \end{minipage}}
  \thanks{%
    Manuscript submitted in September 2015. This work is based on research supported by the U.S. Office of Naval Research (ONR) via grant  \#N00014-10-1-0140, and the U.S. National Science Foundation (NSF) via grant \#1343430.} 
  \thanks{%
    RD, KP, and MNM are with the Department of Electrical and Computer Engineering of the University of Miami, Coral Gables, Florida (e-mail: \{ranga, kamal, mmurthi\}@miami.edu); 
    DS is with the Department of Computer Scienceof the University of Miami, Coral Gables, Florida (e-mail: sarkar@miami.edu).}}

\markboth{%
  IEEE Transactions on Signal and Information Processing Over Networks}{%
  Dabarera, \MakeLowercase{\emph{et al.}}: Consensus in the Presence of Multiple Opinion Leaders ...}

\IEEEpubid{0000--0000/00\$00.00~\copyright~2016 IEEE}
\maketitle


\begin{abstract}
The problem of analyzing the performance of networked agents exchanging evidence   in a dynamic network has recently grown in importance. This problem has relevance in signal and data fusion network applications and in studying opinion and consensus dynamics in social networks. Due to its capability of handling a wider variety of uncertainties and ambiguities associated with evidence, we use the framework of Dempster-Shafer (DS) theory to capture the opinion of an agent. We then examine the consensus among agents in dynamic networks in which an agent can utilize either a \emph{cautious} or \emph{receptive} updating strategy. In particular, we examine the case of \emph{bounded confidence updating} where an agent exchanges its opinion only with neighboring nodes possessing `similar' evidence. In a fusion network, this captures the case in which nodes only update their state based on evidence consistent with the node's own evidence. In opinion dynamics, this captures the notions of Social Judgment Theory (SJT) in which agents update their opinions only with other agents possessing opinions closer to their own. Focusing on the two special DS theoretic cases where an agent state is modeled as a Dirichlet body of evidence and a probability mass function (p.m.f.), we utilize results from matrix theory, graph theory, and networks to prove the existence of consensus agent states in several time-varying network cases of interest. For example, we show the existence of a consensus in which a subset of network nodes achieves a consensus that is adopted by follower network nodes. Of particular interest is the case of multiple opinion leaders, where we show that the agents do not reach a consensus in general, but rather converge to `opinion clusters'. Simulation results are provided to illustrate the main results.
\end{abstract}


\IEEEpeerreviewmaketitle


\section{Introduction}


\tb{\emph{Motivation.}} 
In multi-agent systems like social networks or a group of mobile robots completing a task, agent behavior is determined by local interactions between neighboring agents. In distributed control problems, reaching a consensus state is often necessary for the agents to achieve a control objective \cite{Olfati-Saber2007, Qu2008}. Within the context of distributed sensor fusion, the agents exchange evidence to eventually have a good estimate of unknown parameters  \cite{Xio05}. Agents in a social network typically exchange opinions in some manner with their neighbors, and update their own opinions, possibly resulting in consensus. The work in \cite{Chamley2013} discusses models of opinion diffusion among rational agents and social learning, explaining how opinions emerge or fail to emerge in contexts of economics and social sciences. Understanding the conditions under which a group of agents will reach a possible consensus in a network with time-varying links is a challenging problem that has recently garnered much interest.

\pubidadjcol

\tb{\emph{Previous Work.}} 
Previous work within signal processing, control, and data fusion geared towards analyzing agent consensus formation has mostly modeled an agent state as a real-valued vector \cite{Fortunato2005c, Ren2005, Xio05, Olfati-Saber2007, Qu2008} (see also \cite{Wickramarathne2014JoSTSP} and references therein). While such an assumption is useful in the context of agents achieving consensus on a control vector, or sensing network agents achieving a consensus signal estimate, the assumption of a real-valued vector may not necessarily be always suitable within higher-level fusion, or the analysis of opinion dynamics in social networks. In higher-level fusion, agents could reach a consensus assessment regarding a situation, while in opinion dynamics, agents gravitate to reach a possible consensus opinion. The inherent uncertainty within such applications may necessitate a more structured agent state vector such as a probability mass function (p.m.f.) or an imprecise probabilisitic formalism, such as \emph{Dempster-Shafer (DS) theory} \cite{Shafer1976}. 

Agent states/opinions in complex fusion environments and social networks involve more qualitative and nuanced information and hence are associated with higher and a wider variety of uncertainties. Uncertainty modeling formalisms based on imprecise probability, such as DS theory, provide a convenient framework for such scenarios. The DS theoretic (DST) framework stands apart from alternate uncertainty handling frameworks (such as, fuzzy sets, rough sets, etc.) in that it bears a close relationship to the probabilistic framework \cite{Fag91_UAI}. DST models converge to and/or can be converted to p.m.f. models in limiting cases, thus allowing for a swift generalization of legacy techniques. Moreover, DST models provide a more intuitive and convenient approach for handling the types of uncertainties and the nuances that are characteristic of agent states and opinions in complex fusion environments and social networks \cite{Blackman1999, Wickramarathne2014JoSTSP}. For example. complete ignorance can be  captured via the `vacuous' model and lack of evidence can be captured by allowing evidential support for non-singleton propositions (see Section~\ref{sec:preliminaries}). For the variety of data imperfection types that DST models can conveniently capture, see Table~1 in \cite{Wickramarathne2011ToKDE}. However, much of the work on opinion dynamics in social networks has focused on modeling the agent states as either scalar real numbers, or a vector of real numbers \cite{Lorenz2007}, e.g., see the \emph{Hegselmann-Krause (HK) model} \cite{Krause2000, Hegselmann2002} and the \emph{Deffuant-Weisbuch (DW) model} \cite{Deffuant2000, Weisbuch2004}.

In our recent work \cite{Wickramarathne2014JoSTSP, Dabarera2014Fusion}, we utilized the DST framework to address consensus formation in asynchronous dynamic/ad-hoc networks with applications to high-level fusion networks, and opinion dynamics in social networks. By utilizing DS theory, the methods by which an agent models an opinion and updates its opinion are now equipped with a powerful mechanism for grappling with the uncertainty inherent in the problem, whether in the form of vague agent opinions, or imprecise data in a fusion network (e.g., vague witness statements). In  \cite{Wickramarathne2014JoSTSP}, we presented the foundations of analyzing opinion dynamics and revision of agent state beliefs in asynchronous dynamic/ad-hoc, demonstrating conditions under which consensus is achieved. 

Our more recent work in \cite{Dabarera2014Fusion} utilized ideas from psychology, namely Social Judgment Theory (SJT) \cite{Sherif1980social}, which examines the basic psychological processes underlying the expression of attitudes and their modifiability through communication. In particular, in  \cite{Dabarera2014Fusion}, we specifically used the notion of \emph{bounded confidence} which stems from the concept of \emph{latitude of acceptance} in SJT. In a social network, agents may only communicate and exchange opinions with their neighbors who have similar opinions on a particular topic. In other words, an agent may be willing to update its opinion with the neighboring agent's opinion only if the `distance' to that opinion is within a certain bound of confidence. The opinion exchange models in the HK and DW models account for these bounded confidence notions. The work in \cite{Rad2014} addresses statistical estimation of the bound of confidence. In \cite{Varshney2013} agents are treated as Bayesian decision-makers and Bayes' risk error has been used to estimate the bounds. The bounded confidence assumption is also useful within high-level fusion networks to capture situations where agents may only exchange evidence with agents that have states similar to their own and perhaps avoiding the use of faulty outlier sensors/agents. 

\tb{\emph{Contributions.}} 
As in \cite{Wickramarathne2014JoSTSP, Dabarera2014Fusion}, this present work of ours on agent consensus formation in dynamic networks is formulated within the DST framework. This allows us to embrace a wider variety of uncertainties and nuances that are more representative of agent states/opinions in complex fusion networks and social networks. 

Taking this DST setting as a springboard, for the main contributions of this work, we focus on two special cases of DST agent opinion models, viz., opinions modeled as a p.m.f. and a Dirichlet body of evidence (which allows one to retain DS theory's ability to capture complete ignorance with only a slight increase in computational complexity) \cite{Josang2010}. Utilizing results from matrix theory, graph theory, and networks, we provide a thorough theoretical analysis of the conditions for consensus formation among a group of agents residing within a dynamic network (i.e., the network connectivity graph is dynamic, so that links can appear, or disappear, meaning that the agents may not have the same neighbors at each time instant). Of particular value are the following:

  \tb{(1)}~Each agent may employ either \emph{cautious} or \emph{receptive} opinion update strategies, which essentially determines whether the agent acts as an opinion `leader' or an opinion `follower.' 
  
  \tb{(2)}~The agents employ bounded confidence notions so that an agent exchanges and updates its state/opinion with only those neighbors whose opinions are similar (as measured by a suitable norm) to its own opinion. 
  
  \tb{(3)}~The notion of \emph{opinion dynamic chains} and corresponding results being presented in this work capture network topologies, and conditions under which, where one or two groups of agents interact with another group of agents to drive them toward a consensus opinion or opinion clusters. 
  
  \tb{(4)}~Focusing on the two cases of p.m.f. and DST Dirichlet agent opinion models, the behavior of the network in the presence of multiple opinion leaders is explored. The analytical results require significantly less restrictive assumptions than those in \cite{Dabarera2014Fusion}, which, to the authors' knowledge, is the only other work that deals with multiple opinion leaders. 
  
  \tb{(5)}~These general analytical results are then used to study how opinion leaders interact with opinion followers. In particular, it is shown that the arrival at a consensus opinion or opinion clusters is dependent on the the number of opinion leaders present and the agents' bounds of confidence. 
  
  \tb{(6)}~Several cases of interest in both opinion dynamics and fusion domains are studied. This includes the the \emph{leader-follower problem} \cite{Lu2013} where a subset of the networked agents achieve an opinion cluster which is then  followed and adopted by the remaining agents. 
  
  \tb{(7)}~Finally, experimental evaluations are carried out to explore whether these results on p.m.f. and Dirichlet agent opinions apply to and are valid for agent opinions represented via general DST models.

As elaborated upon earlier, DST framework offers a more intuitive and convenient avenue to capture the types of uncertainties and the nuances associated with agent opinions. This current work and the results regarding p.m.f. and Dirichlet agent opinions constitute an important step toward understanding how agents whose states/opinions are captured via more general DST models behave in terms consensus and opinion cluster formation. 

\tb{\emph{Organization of Paper.}} 
The paper is organized as follows: 
  Section~\ref{sec:preliminaries} covers basic background material. 
  Section~\ref{sec:ourModel} presents our opinion update model with the conditional update equation (CUE) as the governing equation. 
  Our main theoretical contribution on opinion dynamic chains driven by one group (\emph{1-ODC}) and opinion dynamic chains driven by two groups (\emph{2-ODC}) are in Section~\ref{sec:ODC} where we focus on two special cases of the DST agent opinion model: Section~\ref{subsec:PMF} deals with agent opinions captured via p.m.f.s which would appeal to readers with background knowledge in probabilistic opinion representation; Section~\ref{subsec:DirichletBBA} deals with DST Dirichlet agent opinions. Section~\ref{sec:Results} provides experimental validation of our results, including their applicability in scenarios where agent opinions are captured via more general DST models. 


\section{Preliminaries}
\label{sec:preliminaries}


We use $\mathbb{N}$ and $\mathbb{R}$ to denote the integers and reals, respectively. Subscript $(\centerdot)_{\geq 0}$ attached to these are their non-negative counterparts; $\mathbb{R}_{[0,1]}$ denotes the reals taking values in $[0, 1]$. The superscripts $(\centerdot)^N$ and $(\centerdot)^{M\times N}$ denote $N$-sized vector and $(M\times N)$-sized matrix counterparts. For $X=\{X_{ij}\}\in\mathbb{R}^{M\times N}$, $\Vert X\Vert$ denotes its $\infty$-norm, i.e., $\Vert X\Vert=\max_{i\in\ol{1,M}} \sum_{j=1}^N |X_{ij}|$. We use $X>0$ and $X\geq 0$ to denote a matrix/vector with positive and non-negative entries, respectively. A matrix whose entries are all $0$\,s (and is of compatible size) is denoted by $\mb{0}$; an $N$-element vector whose elements are all $1$\,s is denoted by $\mb{1}_N$.

We use $|\Theta|$ to denote the cardinality of set $\Theta$.

\tb{\emph{Relevant Notions from DS Theory.}} 
In DS theory, the \textit{frame of discernment (FoD)} refers to the set $\Theta = \{\theta_1, \cdots, \theta_M\}$ consisting of the mutually exclusive and exhaustive `singleton' propositions $\theta_i$ \cite{Shafer1976}. A singleton proposition $\theta_i\in\Theta$ represents the lowest level of discernible information; $|\Theta|=M$ is the number of independent singleton propositions in the FoD $\Theta$. The power set of the FoD,  $2^{\Theta} = \{A : A \subseteq \Theta \}$,  denotes all the possible subsets of $\Theta$. For $A\subseteq\Theta$, $\ol{A}$ denotes all singletons in $\Theta$ that are not in $A$.

\emph{Basic Belief Assignment (BBA).} 
A \emph{basic belief assignment (BBA)} or \textit{mass assignment} is a mapping $m(\cdot): 2^{\Theta} \mapsto[0, 1]$ such that $\sum_{A \subseteq \Theta} m(A)=1$ and $m(\emptyset)=0$. The BBA measures the ``support" assigned to proposition $A \subseteq \Theta$. Propositions that receive non-zero mass are referred to as \emph{focal elements.} The set of focal elements is the \emph{core $\mc{F}$.} The triplet $\mc{E}=\{\Theta, \mc{F}, m\}$ is referred to as the \textit{body of evidence (BoE).} The \emph{mass vector} corresponding to the BoE $\mc{E}$ is 
\begin{multline}
  \hspace*{-0.1in}
  \tb{m}
    =\left[
       m(\emptyset), m(\theta_1), \ldots, m(\theta_M), 
       m(\theta_1\theta_2), \ldots, m(\theta_1\theta_M),
     \right. \\
     \left.
       m(\theta_1\theta_2\theta_3), \ldots, 
       m(\theta_1\theta_2\theta_M), \ldots, m(\Theta)
     \right]
     \in\mathbb{R}_{[0,1]}^{2^M}.\hspace*{-0.1in}
\end{multline}

Note that, masses are assigned to all propositions in $2^{\Theta}$. DS theory captures the notion of \emph{ignorance} by allowing composite (i.e., non-singleton) propositions to be focal elements, e.g., the mass assignment $m(\theta_i\theta_j)>0$ represents ignorance or lack of evidence to differentiate between the singletons $\theta_i$ and $\theta_j$. The \textit{vacuous BBA} which $\Theta$ as its only focal element (so that $m(\Theta)=1$) captures complete ignorance.

\emph{Belief and Plausibility.} 
Given a BoE, $\mc{E}\equiv\{\Theta, \mc{F}, m\}$, the \textit{belief} $\mathrm{Bl}: 2^\Theta\mapsto [0, 1]$, defined as $\mathrm{Bl}(A)=\sum_{B\subseteq A} m(B)$, represents the total support committed to $A$ without also being committed to its complement $\ol{A}$. The \textit{plausibility} $\mathrm{Pl}: 2^\Theta\mapsto [0, 1]$, defined as $\mathrm{Pl}(A)=1-\mathrm{Bl}(\ol{A})$, corresponds to the total belief that does not contradict $A$. The \textit{uncertainty} of $A$ is $\mathrm{Un}(A)=[\mathrm{Bl}(A), \mathrm{Pl}(A)]$. We also use the notation $\widehat{\mc{F}}=\{A\subseteq\Theta: \mathrm{Bl}(A)>0\}$.

A BoE is called \emph{Bayesian} if each focal element is a singleton. For a Bayesian BoE, the BBA, belief, and plausibility, all reduce to the same probability (i.e., p.m.f.) assignment. The expressive power that the DST framework wields a high cost: For a given FoD $\Theta$, where $|\Theta|=M$, a DST model allocates $2^M-2$ mass assignments; only $M-1$ probability assignments are required for a p.m.f. Much advances have been made for the purpose of mitigating the associated computational burden \cite{Bauer1997IJAR, Wilson2000_chap, Wic10_SMCB_cct}. A special DST model which retains the ability to capture complete ignorance with only a slight increase in computational complexity is the \emph{Dirichlet BoE} (so named because of its close relationship with Dirichlet probability distributions \cite{Josang2010}). The singletons $\{\theta_i\}$ and $\Theta$ constitute the only focal elements of a Dirichlet BoE, thus requiring only $M$ mass assignments.

\emph{DST Conditionals.} 
Of the various notions of DST conditionals that abound in the literature, the Fagin-Halpern (FH) conditional offers a unique probabilistic interpretation and constitutes  a natural transition to the Bayesian conditional notion  \cite{Fag91_UAI, Wic10_SMCB_cct}. The extensive study in \cite{Den94} identifies several attractive properties of the FH conditionals including its equivalence to other popular notions of DST conditionals. 

\begin{mydef}[FH Conditionals]
For the BoE $\mc{E}=\{\Theta, \mc{F}, m\}$ and $A\subseteq\Theta$ s.t. $A\in\widehat{\mc{F}}$, the \emph{conditional belief} $\mathrm{Bl}(B|A): 2^\Theta\mapsto [0, 1]$ and the \emph{conditional plausibility} $\mathrm{Pl}(B|A): 2^\Theta\mapsto [0, 1]$ of $B$ given $A$ are, respectively, 
\begin{align*}
  \mathrm{Bl}(B|A)	
    &=\mathrm{Bl}(A\cap B)
        /[\mathrm{Bl}(A\cap B)+\mathrm{Pl}(A\cap\ol{B})];
      \notag \\
  \mathrm{Pl}(B|A)	
    &=\mathrm{Pl}(A\cap B)
        /[\mathrm{Pl}(A\cap B)+\mathrm{Bl}(A\cap\ol{B})].
  \tag*{\QEDclosed}
\end{align*}
\end{mydef}

The conditional core theorem \cite{Wic10_SMCB_cct} can be utilized to directly identify the conditional focal elements to improve computational performance when applying FH conditionals.

\emph{DST Distance Measure.} 
In our work, we need a distance measure which captures the closeness between DST BoEs. Among possible alternatives that have appeared in the literature, we use the DST distance measure in \cite{Jousselme2001, Martin2008}. 

\begin{mydef}[Distance Between BoEs] \cite{Jousselme2001} 
\label{def:distance}
The distance between the two agent BoEs $\mc{E}_i=\{\Theta, \mc{F}_i, m_i\}$ and $\mc{E}_j=\{\Theta, \mc{F}_j, m_j\}$ is 
\[
  \Vert\mc{E}_i-\mc{E}_j\Vert_J
    =\left[
       0.5\,(\tb{m}_i-\tb{m}_j)^{^T} 
       \mc{D}\,
       (\tb{m}_i-\tb{m}_j)
     \right]^{1/2}
    \in\mathbb{R}_{[0,1]},
\]
where $\tb{m}_i, \tb{m}_j\in\mathbb{R}_{\geq 0}^{2^M}$ are the mass vectors associated with the BoEs $\mc{E}_i$ and $\mc{E}_j$, respectively; $\mc{D}=\{d_{mn}\}\in\mathbb{R}_{\geq 0}^{2^M\times 2^M}$, with $d_{mn}=|A_m\cap A_n|/|A_m\cup A_n|,\,A_m, A_n\subseteq\Theta$, with $|\emptyset\cap\emptyset|/|\emptyset\cup\emptyset|\equiv 0$.
\hspace*{\fill}\QEDclosed
\end{mydef}

\tb{\emph{Relevant Notions from Graph Theory.}} 
\cite{Newman2012N}
We use $\mc{G}_k=(V, E_k)$ to denote a directed graph at discrete-time (DT) instant $k\in\mathbb{N}_{\geq 0}$. Here, $e_{ij}\in E_k$ represents a unidirectional edge from node $V_j\in V$ to node $V_i\in V$. We use $A_k$ to identify the $(N\times N)$ adjacency matrix associated with $E_k$. 

Consider the directed graph $\mc{G}_k=(V, E_k)$. The \emph{out-component of vertex $V_i\in V$} is the set of vertices (including vertex $V_i$ itself) reachable via directed paths from vertex $V_i$. The \emph{in-component of vertex $V_i\in V$} is the set of vertices (including vertex $V_i$ itself) from which vertex $V_i$ is reachable via directed paths. 

\tb{\emph{Relevant Notions from Stochastic Matrix Theory.}} We say that $X\in\mathbb{R}_{[0,1]}^{N\times N}$ is \emph{stochastic} if $\sum_{j=1}^N X_{ij}=1,\,\forall i\in\ol{1,N}$ and $\Vert X\Vert=1$; we say that $X\in\mathbb{R}_{[0,1]}^{N\times N}$ is \emph{sub-stochastic} if $\sum_{j=1}^N X_{ij}\leq 1,\,\forall i\in\ol{1,N}$, and $\exists\,i\in\ol{1,N}$ s.t. $\sum_{j=1}^N X_{ij}<1$. Stochastic/sub-stochastic matrices and the limiting behavior of their products play a critical role in our work. 


\section{DST Modeling of Opinion Dynamics}
\label{sec:ourModel}


In the work that follows, we consider $N$ agents embedded in the directed graph $\mc{G}_k=(V, E_k)$. Here, for $i\in\ol{1,N}$, the node $V_i\in V$ in $\mc{G}_k$ represents the $i$-th agent and the directed edge $e_{ij}\in E_k$ represents an unidirectional information exchange link from the $j$-th agent to the $i$-th agent (i.e., the $i$-th agent can receive information from the $j$-th agent). Unless otherwise mentioned, the opinion of the $i$-th agent at time instant $k\in\mathbb{N}_{\geq 0}$ is taken to be captured via the BoE $\mc{E}_{i,k}=\{\Theta, \mc{F}_{i,k}, m_i(\cdot)_k\},\,i\in\ol{1, N}$. We assume that the agent opinion BoEs are associated with the identical FoD $\Theta$. 

\begin{mydef}[Opinion Profile] 
\label{def:OpinionProfile}
Consider the agent BoEs $\mc{E}_{i,k}=\{\Theta, \mc{F}_{i,k}, m_i(\cdot)_k\},\,i\in\ol{1, N},\,k\in\mathbb{N}_{\geq 0}$. The \emph{opinion profile} of $B\subseteq\Theta$ at $k\in\mathbb{R}_{\geq 0}$ is 
\[
  \pmb{\pi}(B)_k
    =[m_1(B)_k, \ldots, m_N(B)]^T
     \in\mathbb{R}_{[0,1]}^N,
\]
with $\pmb{\pi}(B)_0\in\mathbb{R}_{[0,1]}^N$ denoting its \emph{initial state.} 
\hspace*{\fill}\QEDclosed
\end{mydef}

  
\subsection{Bounded Confidence}


For each agent, define the following sets of neighborhood agents at time instant $k$:
\begin{alignat}{2}
  &\mc{N}_{i,k} 
    &
      &=\left\{
          V_j\in V: 
          j\in\ol{1,N},\textrm{ and }e_{ij}\in E_k
        \right\};
        \notag \\
  &\mc{N}_{i,k}(\varepsilon_i) 
    &
      &=\left\{
          V_j\in\mc{N}_{i,k}: 
          \Vert\mc{E}_{i,k}-\mc{E}_{j,k}\Vert_J
            \leq\varepsilon_i
        \right\},
\end{alignat}
where $\Vert\cdot\Vert_J$ refers to the distance measure in Definition~\ref{def:distance} (while any valid norm applicable for DST BoEs could be used); $\varepsilon_i\geq 0$ is the \emph{latitude of acceptance} or \emph{bound of confidence} associated with the $i$-th agent. So, $\mc{N}_{i,k}(\varepsilon_i)$ denotes the neighbors of the $i$-th agent at time $k$ left after `pruning' the links subjected to the bound of confidence requirement. With $\pmb{\varepsilon}=[\varepsilon_1, \ldots, \varepsilon_N]^T$, let
\begin{equation}
  \label{eq:Gdot}
  \mc{G}_k^{\dag}(\pmb{\varepsilon}) 
    =(V, E_k^{\dag}(\pmb{\varepsilon})),
\end{equation}
where $E_k^{\dag}(\pmb{\varepsilon})=\left\{e_{ij}\in E_k: \Vert\mc{E}_{i,k}-\mc{E}_{j,k}\Vert_J\leq\varepsilon_i\right\}$.

The bounded confidence process of updating an agent's opinion is as follows  \cite{Fortunato2005c, Lorenz2007}: the $i$-th agent updates its BoE $\mc{E}_i$ in response to the opinion BoE $\mc{E}_j$ of its neighbor,  the $j$-th agent, only if $j\in\mc{N}_i(\varepsilon_i)$. In \cite{Deffuant2000}, the threshold $\varepsilon_i$ is referred to as an \emph{openness character.}  Another interpretation views $\varepsilon_i$ as an \emph{uncertainty,} i.e., if the $i$-th agent possesses an opinion with some degree of uncertainty $\varepsilon_i$, then it ignores the views of those neighbor agents who fall outside its uncertainty range. 


\subsection{Opinion Updating and Consensus Formation}
\label{subsec:linear_Algebraic_Proof}


\tb{\emph{Opinion Updating.}} 
In what follows, we will use $\mc{S}$ to identify the indices corresponding to a subset of the agents in $V$, i.e., $\mc{S}\subseteq\{1, 2, \ldots, N\}$ s.t. $V_i,\,\forall i\in\mc{S}$, identifies a subset of agents in $V$. To proceed, we adopt the following 

\begin{mydef}[Opinion Clusters, Consensus] 
\label{def:consensus}
Let $\mc{E}_{i,k},\,i\in\ol{1,N},\,k\in\mathbb{N}_{\geq 0}$, denote the opinions of $N$ agents embedded within the network $\mc{G}_k=(V, E_k)$, where each agent repeatedly updates its state by iterative opinion exchange. Let $\mc{S}\subseteq\{1, 2, \ldots, N\}$ identify a subset of the agents in $V$.

  \tb{(i)}~Suppose $\lim_{k\to\infty} \Vert\mc{E}_{i,k}-\mc{E}_{j,k}\Vert_J=0,\,\forall i,j\in\mc{S}$ (or equivalently, $\lim_{k\to\infty} \mc{E}_{i,k}\equiv\mc{E}_{*},\,\forall i\in\mc{S}$) and suppose $\lim_{k\to\infty} \mc{E}_{i,k}\neq\mc{E}_{*},\,\forall i\in\ol{1,N}\setminus\mc{S}$. Then, the agents in $\mc{S}$ are said to form an \emph{opinion cluster.}
  
  \tb{(ii)}~The agents are said to reach a \emph{consensus} if $\mc{S}=V$, i.e., all the agents in $V$ form a single opinion cluster. 
\hspace*{\fill}\QEDclosed
\end{mydef}

Henceforth, our results will only be stated with the formation of a consensus in mind (e.g., see Lemma~\ref{lem:consensus_eta1}). Due to (ii) above, these results can easily be reformulated so that they pertain to the formation of opinion clusters. 

In our work, we assume that each agent updates its opinion in accordance with the conditional update equation (CUE):

\begin{mydef}[Conditional Update Equation (CUE)] 
\label{def:CUEm}
\cite{Premaratne2007a, Pre09_ICIF, Wickramarathne2014JoSTSP} 
Suppose the $i$-th agent updates its opinion $\mc{E}_i$ by taking into account its neighboring agents $j\in\mc{N}_{i,k}(\varepsilon_i)$, where $\varepsilon_i>0$ is the $i$-th agent's bound of confidence. The CUE-based updated opinion of the $i$-th agent is 
\[
  \mathrm{Bl}_i(B)_{k+1} 
    =\alpha_{i,k}\mathrm{Bl}_i(B)_k
       +\!\!
        \sum_{j\in\mathbb{I}_{i,k}\setminus i} 
        \sum_{A\in\mc{\hat{F}}_{j,k}} 
        \beta_{ij}(A)_k\mathrm{Bl}_j(B|A)_k.
\]
Here, the index set $\mathbb{I}_{i,k}$ identifies the agents in $\mc{N}_{i,k}(\varepsilon_i)$; the CUE parameters $\alpha_{i,k}, \beta_{ij}(\cdot)_k\in\mathbb{R}_{[0,1]}$ satisfy
\[
  \alpha_{i,k}
    +\sum_{j\in\mathbb{I}_{i,k}\setminus i} 
     \sum_{A\in\mc{\hat{F}}_{j,k}} \beta_{ij}(A)_k 
    =1.
  \tag*{\QEDclosed}
\]
\end{mydef}

The above CUE can also be expressed directly in terms of DST mass values. Alternately, the updated mass values can be computed from the updated belief values. Note that, with agents utilizing CUE-based opinion updating, the edge `weights' along both directions of an edge are not equal in general because the CUE weights $\beta_{ij}(\cdot)$ and $\beta_{ji}(\cdot)$ are not necessarily the same. It is for this reason that we treat the underlying graph $\mc{G}_k$ as directed. 

The work in \cite{Premaratne2007a, Pre09_ICIF, Wickramarathne2014JoSTSP} also provides various strategies for selecting the CUE parameters. Of particular importance are two particular strategies:

\begin{mydef}[Receptive and Cautious Update Strategies]
\label{def:parameters}
Consider the $i$-th agent updating its opinion $\mc{E}_i$ according to the CUE-based update in Definition~\ref{def:CUEm}. 

  \tb{(i)}~The $i$-th agent is said to employ \emph{receptive updating} if $\beta_{i,j}(A)_k\propto m_j(A)_k$, and receptively updating agents are referred to as \emph{opinion leaders.} 

  \tb{(ii)}~The $i$-th agent is said to employ \emph{cautious updating} if $\beta_{ij}(A)_k\propto  m_i(A)_k$, and cautiously updating agents are referred to as \emph{opinion followers.} 
\hspace*{\fill}\QEDclosed
\end{mydef}

Receptive updating `weighs' the incoming evidence according to the support each focal element receives from the incoming BoEs. This has an interesting Bayesian interpretation: it reduces to a weighted average of the p.m.f.s of the BoEs. Cautious updating `weighs' the incoming evidence according to the support each focal elements receives from the BoE being updated. We consider three cases:

\tb{(1)}~\emph{No Opinion Leaders:} 
This is the most common scenario that appears in the literature \cite{Deffuant2000, Hegselmann2002, Lorenz2007}. Here, all agents are receptively updating and no opinion leaders are present. 

\tb{(2)}~\emph{Single Opinion Leader:} 
Here, all agents employ receptive updating except one opinion leader. This is the scenario considered in typical leader-follower models \cite{Ren2005}, and the recent work in \cite{Wickramarathne2014JoSTSP}).

\tb{(3)}~\emph{Multiple Opinion Leaders:} 
Here, there are multiple cautiously updating opinion leaders, generally with different initial opinions. To our knowledge, this case has not been addressed prior to the work in \cite{Dabarera2014Fusion} which was based on the work in \cite{Krause2000, Dittmer2001}. It provides sufficient conditions for consensus/cluster formation under certain strong assumptions (e.g., the agent opinions can be ordered in what is referred to as an $\pmb{\varepsilon}$-chain). We will relax such assumptions by utilizing properties of products of stochastic/sub-stochastic matrices. 


\section{Opinion Dynamic Chains with Opinion Leaders}
\label{sec:ODC}


While the DST framework provides a powerful tool for grappling with the types of uncertainties and nuanced information prevalent in complex fusion and social networks, the expressive power inherent in DST models levy a high computational burden. A Dirichlet BoE \cite{Josang2010} is a special type of DST BoE which retains DS theory's ability to capture complete ignorance with only a slight increase in computational effort. 
 
With our DST opinion model in place, we now focus on probabilistic (i.e., p.m.f.) and DST Dirichlet agent opinions. 


\subsection{Probabilisitic Agent Opinions}
\label{subsec:PMF}


With the DST BoEs $\mc{E}_{i,k},\,i\in\ol{1, N},\,k\in\mathbb{N}_{\geq 0}$, possessing only singleton focal elements, we have probabilistic agent opinions. In this case, the CUE-based opinion update in Definition~\ref{def:CUEm} reduces to the DT dynamic system
\begin{equation}
  \label{eq:DTDsys}
  \pmb{\pi}(\theta_p)_{k+1} 
    =W_k\,\pmb{\pi}(\theta_p)_k,\;
     p\in\ol{1,M},
\end{equation}
where $\pmb{\pi}(\centerdot)_k\in\mathbb{R}_{[0,1]}^N$, $W_k=\{w_{ij,k}\}\in\mathbb{R}_{[0,1]}^{N\times N}$ is row-stochastic \cite{Hartfiel2002}. When all agents employ receptive updating, 
\begin{equation}
  \label{eq:wijk}
   w_{ij,k}
     =\begin{cases}
        \alpha_{i,k},
          & \text{for $i=j$}; \\
        (1-\alpha_{i,k})/|\mc{N}_{i,k}(\varepsilon_i)|,
          & \text{for $j\in\mc{N}_{i,k}(\varepsilon_i)$}; \\
        0,
          & \text{otherwise}.
      \end{cases}
\end{equation} 
As in \cite{Lorenz2006}, we refer to $W_k$ as the \emph{confidence matrix} because $w_{ij,k}$ represents the weight the $i$-th agent attaches to the opinion of the $j$-th agent at time step $k$. Note that, $W_k$ constitutes the weighted adjacency matrix of $\mc{G}_k^{\dag}(\pmb{\varepsilon})$ in \eqref{eq:Gdot}.

We proceed with 

\begin{myprop}
\label{prop:cautious}
In an environment where agents possess probabilistic agent opinions, the opinion of a cautiously updating agent is invariant.
\hspace*{\fill}\QEDopen
\end{myprop}

\emph{Proof.}
Suppose the $i$-th agent employs a cautiously updating strategy. Its CUE parameters satisfy (see Definition~\ref{def:parameters})
\[
  \beta_{ij}(B)_k
    =\mu_{ij,k}m_i(A)_k;\;\;
  \alpha_{i,k}
    +\sum_{j\neq i} \sum_{A\in\mc{\hat{F}}_{j,k}} 
     \mu_{ij,k} 
    =1,
\]
for singletons $A, B\in\Theta$. But, for singleton propositions $A, B\in\Theta$, $\mathrm{Bl}(B|A)=1$ only if $B=A$, and $\mathrm{Bl}(B|A)=0$ otherwise \cite{Wic10_SMCB_cct}. Using this and the fact that for singletons propositions belief and masses are equal, the CUE-based update in Definition~\ref{def:CUEm} for the $i$-th agent reduces to $m_i(B)_{k+1}=m_i(B)_k,\,\forall k\in\mathbb{N}_{\geq 0}$.
\hspace*{\fill}\QEDclosed

\begin{mydef}[Left (or Backward) Products]
\cite{Daubechies1992}
\label{def:bwprod}

  \tb{(i)}~\emph{Left Product:} The \emph{left product} of the sequence of matrices $\{W_k\},\,W_k\in\mathbb{R}^{N\times N}$, is
\[
  W_{k:\ell}
    =\begin{cases}
       I,
         & \text{for $k<\ell$}; \\
       W_{\ell},
         & \text{for $k=\ell$}; \\
       W_kW_{k-1}\cdots W_{\ell},
         & \text{for $k>\ell$}.
     \end{cases}
\]

  \tb{(ii)}~\emph{Left-Converging Product:} The left product $W_{k:0}$ is said to be \emph{left-converging} if $\lim_{k\to\infty} W_{k:0}$ exists, in which case we write $W_{\infty}=\lim_{k\to\infty} W_{k:0}$. 
\hspace*{\fill}\QEDclosed
\end{mydef}

Note that the dynamic system in \eqref{eq:DTDsys} can be expressed as $\pmb{\pi}(\theta_p)_{k+1}=W_{k:0}\,\pmb{\pi}(\theta_p)_0$. Thus, whenever $W_{\infty}$ exists, 
\begin{equation}
  \label{eq:DTDsys3}
  \lim_{k\to\infty} \pmb{\pi}(\theta_p)_{k+1} 
    =W_{\infty}\,\pmb{\pi}(\theta_p)_0.
\end{equation}
Clearly, the convergence of the agent opinions depends on the existence and the nature of $W_{\infty}$. When $W_{\infty}$ exists, let us denote the converged opinion profile for $\theta_p$ as $\pmb{\pi}_{*}(\theta_p)\in\mathbb{R}_{[0,1]}^N$. Consensus (in the sense of Definition~\ref{def:consensus}) is a special case of a converged opinion profile. 

\begin{mylem}
\label{lem:consensus_eta1}
The agents form a  consensus iff $\exists\,\pmb{\eta}=\{\eta_p\}\in\mathbb{R}_{[0,1]}^M$ s.t., $\pmb{\pi}_{*}(\theta_p)=\eta_p\mb{1}_N,\forall p\in\ol{1,M},\,\forall i\in\ol{1,N}$. 
\hspace*{\fill}\QEDopen
\end{mylem}

\emph{Proof.}
Suppose $\pmb{\pi}_{*}(\theta_p)=\eta_p\mb{1}_N,\forall p\in\ol{1,M},\,\forall i\in\ol{1,N}$, for some $\pmb{\eta}=\{\eta_p\}\in\mathbb{R}_{[0,1]}^M$. This clearly means that $\lim_{k\to\infty} \mc{E}_{i,k}\equiv\mc{E}_{*},\,\forall i\in\ol{1,N}$. Thus, the agents form a consensus. Conversely, if $\lim_{k\to\infty} \mc{E}_{i,k}\equiv\mc{E}_{*},\,\forall i\in\ol{1,N}$, we must clearly have $\pmb{\pi}_{*}(\theta_p)=\eta_p\mb{1}_N,\forall p\in\ol{1,M},\,\forall i\in\ol{1,N}$.
\hspace*{\fill}\QEDclosed

The limiting behavior of the stochastic matrix product $\{W_k\}$ plays a crucial role in consensus analysis when DST BoEs possess only singleton focal elements. 

\begin{mylem}
\label{lem:consensus_evt}
Consider the stochastic chain $\{W_k\},\,k\in\mathbb{N}_{\geq 0}$, s.t. $W_{\infty}=\mb{1}\,\mb{v}^T$ for some stochastic vector $\mb{v}\in\mathbb{R}_{[0,1]}^N$. Then, the agents reach the consensus $\pmb{\pi}_{*}(\theta_p)=(\mb{v}^T\pmb{\pi}(\theta_p)_0)\,\mb{1}$, where $\pmb{\pi}(\theta_p)_0\in\mathbb{R}_{[0,1]}^N$ denotes the initial opinion profile.
\hspace*{\fill}\QEDopen
\end{mylem}

\emph{Proof.}
Use \eqref{eq:DTDsys3}: $ \lim_{k\to\infty} \pmb{\pi}(\theta_p)_{k+1}=\mb{1}\,\mb{v}^T\pmb{\pi}(\theta_p)_0=\eta_p\mb{1}$, where $\eta_p=\mb{v}^T\pmb{\pi}(\theta_p)_0$. So, from Lemma~\ref{lem:consensus_eta1}, we achieve a consensus. 
\hspace*{\fill}\QEDclosed

\subsubsection{No Opinion Leaders}
\label{subsec:0}

This is the most widely studied scenario and many consensus-related results applicable to this case are available \cite{Chatterjee1977, Ren2005, Lorenz2006, Lorenz2007, Touri_2014}. The work in \cite{Lorenz2006} studied convergence when each element in the stochastic chain $\{W_k\}$ has positive diagonals. From a graph theoretic viewpoint, this implies that there is a path from each vertex to itself; from an opinion dynamics perspective, this is referred to as having the \emph{self-communicating} property \cite{Lorenz2006}. 

\smallskip
\subsubsection{Single Opinion Leader}
\label{subsec:1}

We first introduce 

\begin{mydef}[Opinion Dynamics Chain Driven By One Group (1-ODC)] 
\label{def:1-ODC} 
The directed dynamic network $\mc{G}_k^{\dag}(\pmb{\varepsilon})=(V, E_k^{\dag}(\pmb{\varepsilon}))$ in \eqref{eq:Gdot} is said to be an \emph{opinion dynamics chain driven by one group (1-ODC)} if its corresponding confidence matrix $W_k$ in \eqref{eq:DTDsys} can be expressed as the lower block triangular matrix 
\[
  W_k
    =\begin{bmatrix}
       A_k & \mb{0} \\
       C_k & D_k
     \end{bmatrix},
\]
where $A_k\in\mathbb{R}_{[0,1]}^{N_C\times N_C}$ and $D_k\in\mathbb{R}_{[0,1]}^{N_{out}\times N_{out}}$, and the other matrices have compatible sizes. 
\hspace*{\fill}\QEDclosed
\end{mydef}

\begin{figure}[ht]
  \centering
  \includegraphics[width=0.30\textwidth]{%
    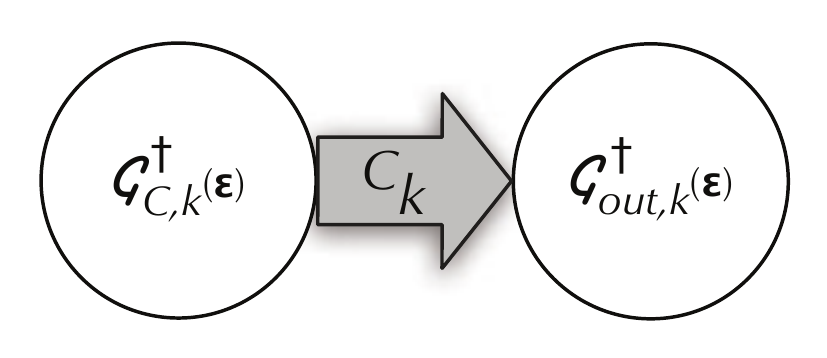}
  \caption{An opinion dynamics chain driven by one group (1-ODC).}
  \label{fig:absorbing_chain}
\end{figure}

One may view an 1-ODC as consisting of a `central' component $\mc{G}_{C,k}^{\dag}(\pmb{\varepsilon})=(V_C, E_{C,k}^{\dag}(\pmb{\varepsilon}))$ (with $N_C$ agents) and another component $\mc{G}_{out,k}^{\dag}(\pmb{\varepsilon})=(V_{out}, E_{out,k}^{\dag}(\pmb{\varepsilon}))$ (with $N_{out}$ agents) s.t. no agent in $\mc{G}_{out,k}^{\dag}(\pmb{\varepsilon})$ belongs to the in-component of $\mc{G}_{C,k}^{\dag}(\pmb{\varepsilon})$, for any $k\in\mathbb{N}_{\geq 0}$. See Fig.~\ref{fig:absorbing_chain}. Note that, $A_k$ and $D_k$ correspond to the confidence matrices of agents in $\mc{G}_{C,k}^{\dag}(\pmb{\varepsilon})$ and $\mc{G}_{out,k}^{\dag}(\pmb{\varepsilon})$, respectively. 

\begin{mythm} 
\label{theorem:absorbing_chain} 
Consider agents embedded in a 1-ODC employing a CUE-based update strategy. Furthermore, suppose that 

  \tb{(a)}~$\lim_{n\to \infty} A_{n:0}=\mb{1}_{N_C}\mb{v}_A^T$, where $\mb{v}_A\in\mathbb{R}_{[0,1]}^{N_C}$ is a stochastic vector so that the agents in $V_C$ reach their own consensus, and 
  
  \tb{(b)}~$\Vert D_k\Vert\leq\rho<1,\,\forall k\in\mathbb{N}_{\geq 0}$. 
  
Then, the agents in $V$ (i.e., agents in $V_C$ and $V_{out}$)  reach a consensus at the consensus reached by the agents in $V_C$. 
\hspace*{\fill}\QEDopen
\end{mythm}

\emph{Proof.}
The opinion update strategy yields the dynamic system
\begin{equation}
  \label{eq:absorbing_chain1}
  \begin{bmatrix}
    \pmb{\pi}_A(\theta_p)_{k+1} \\
    \pmb{\pi}_D(\theta_p)_{k+1}
  \end{bmatrix}
    =W_k
     \begin{bmatrix}
       \pmb{\pi}_A(\theta_p)_k \\
       \pmb{\pi}_D(\theta_p)_k
     \end{bmatrix},
\end{equation}
where $k\in\mathbb{N}_{\geq 0}$, $\theta_p\in\Theta$, and
\begin{equation}
  \label{eq:fnzero}
  W_k
    =\begin{bmatrix}
       A_k & \mb{0} \\
       C_k & D_k
     \end{bmatrix}
  \implies
  W_{n:0}
    =\begin{bmatrix}
       A_{n:0} & \mb{0} \\
       P_n & D_{n:0}
     \end{bmatrix}.
\end{equation}
Here the sub matrices have sizes compatible with $W_k$. Use $W_{n+1:0}=W_{n+1}W_{n:0}$ to get
\begin{equation}
  \label{eq:pnplus1}
  P_{n+1}
    =C_{n+1}A_{n:0}+D_{n+1}P_n,\;
  P_0
    =C_0, 
\end{equation}
for $n\in\mathbb{N}_{\geq 0}$. Due to the row-stochasticity of $W_k$, we have
\begin{equation}
  \mb{1}_{N_{out}}
    =C_k\mb{1}_{N_C}+D_k\mb{1}_{N_{out}},\;
     \forall k\in\mathbb{N}_{\geq 0}.
  \label{eq:row_stoch}
\end{equation}

Subtract $\mb{1}_{N_{out}}\mb{v}_A^T$ from both sides of \eqref{eq:pnplus1}:
\begin{align*}
  &P_{n+1}-\mb{1}_{N_{out}}\mb{v}_A^T 
   \notag \\
  &\quad 
     =C_{n+1}A_{n:0}+D_{n+1}P_n
        -\mb{1}_{N_{out}}\mb{v}_A^T 
   \notag \\
  &\quad 
     =C_{n+1}A_{n:0}+D_{n+1}P_n
        -[C_{n+1}\mb{1}_{N_C}+D_{n+1}\mb{1}_{N_{out}}]
         \,\mb{v}_A^T 
   \notag \\
  &\quad 
     =C_{n+1}[A_{n:0}-\mb{1}_{N_C}\mb{v}_A^T]  
        +D_{n+1}[P_n-\mb{1}_{N_{out}}\mb{v}_A^T],
\end{align*}
for $n\in\mathbb{N}_{\geq 0}$. Here, we have used \eqref{eq:row_stoch}. Employing the notation $\Delta P_n=P_n-\mb{1}_{N_{out}}\mb{v}_A^T,\, n\in\mathbb{N}_{\geq 0}$, we express this as
\[
  \Delta P_{n+1}
    =C_{n+1}[A_{n:0}-\mb{1}_{N_C}\mb{v}_A^T] 
       +D_{n+1}\Delta P_n,\;
     n\in\mathbb{N}_{\geq 0}. 
\]
Then, we may bound $\Vert\Delta P_{n+1}-D_{n+1}\Delta P_n\Vert$ as
\begin{align}
  \label{eq:deltap_12}
  &\vert
     \Vert
       \Delta P_{n+1}\Vert-\Vert D_{n+1}\Delta P_n
     \Vert
   \vert
   \notag \\
  &\quad
     \leq
      \Vert\Delta P_{n+1}-D_{n+1}\Delta P_n\Vert  
        =\Vert 
           C_{n+1}[A_{n:0}-\mb{1}_{N_C}\mb{v}_A^T]
         \Vert 
   \notag \\
  &\quad
     \leq 
      \Vert A_{n:0}-\mb{1}_{N_C}\mb{v}_A^T\Vert.
\end{align}

We proceed by noting that $\lim_{n\to\infty} A_{n:0}=\mb{1}_{N_C}\mb{v}_A^T$, where $\mb{v}_A$ is a $N_C$-sized stochastic vector, implies that the agents in $V_C$ converges to a consensus (see Lemma~\ref{lem:consensus_evt}). Hence, given an arbitrary $\epsilon_A>0$, $\exists N_A\in\mathbb{N}_{\geq 0}$ s.t. 
\begin{equation}
  \label{eq:epsilon_A}
  \Vert A_{n:0}-\mb{1}_{N_C}\mb{v}_A^T\Vert 
    <\epsilon_A,\;
     \forall n\geq N_A. 
\end{equation}
From \eqref{eq:deltap_12} and \eqref{eq:epsilon_A}, we can obtain the following: given an arbitrary $\epsilon_A >0$, $\exists N_A\in\mathbb{N}_{\geq 0}$ s.t.
\[
  \vert
    \Vert\Delta P_{n+1}\Vert
      -\Vert D_{n+1}\Delta P_n\Vert
  \vert 
    <\epsilon_A,\; 
     \forall n\geq N_A. 
\]
So, for $n\geq N_A$, we have 
\[
  \Vert\Delta P_{n+1}\Vert 
    <\Vert D_{n+1}\Delta P_n\Vert+\epsilon_A
    <\rho\,\Vert\Delta P_n\Vert+\epsilon_A,
\]
where we use the fact that $\Vert D_{n+1}\Vert\leq\rho<1,\,\forall n\geq N_A$. This upper bound for $\Vert\Delta P_{n+1}\Vert$ yields
\begin{equation}
  \label{eq:deltapnal}
  \Vert\Delta P_{N_A+L}\Vert 
    <\epsilon_A 
     \sum_{\ell=0}^{L-1} 
     \rho^{\ell}+\rho^L\Vert \Delta P_{N_A}\Vert,\;
     L\geq 0. 
\end{equation} 
We note that 
\[
  \lim_{L\to\infty} 
  \left( 
    \epsilon_A\sum_{\ell=0}^{L-1} \rho^{\ell} 
      +\rho^L\Vert\Delta P_{N_A}\Vert  
  \right) 
    =\frac{\epsilon_A}{1-\rho}, 
\]
i.e., given an arbitrary $\epsilon_P>0$, $\exists L_P\in\mathbb{N}_{\geq 0}$ s.t. 
\[
  \epsilon_A\sum_{\ell=0}^{L-1} \rho^{\ell} 
    +\rho^L\Vert\Delta P_{N_A}\Vert 
    -\frac{\epsilon_A}{1-\rho} 
    <\epsilon_P,\;
     \forall L\geq L_P. 
\]
Use this in \eqref{eq:deltapnal}: $\Vert\Delta P_{N_A+L}\Vert<\epsilon_P+\epsilon_A/(1-\rho),\;\forall L \geq L_P$. In other words, 
\begin{equation}
  \lim_{n\to\infty} \Vert\Delta P_n\Vert 
    =\lim_{n\to\infty} 
     \Vert P_n-\mb{1}_{N_{out}}\mb{v}_A^T\Vert 
    =0. 
    \label{eq:pn_lim}
\end{equation}
Using \eqref{eq:fnzero} in \eqref{eq:absorbing_chain1}, we get
\begin{equation}
  \label{eq:absorbing_chain2}
  \begin{bmatrix}
    \pmb{\pi}_A(\theta_p)_{n+1} \\
    \pmb{\pi}_D(\theta_p)_{n+1}
  \end{bmatrix}
    =\begin{bmatrix}
       A_{n:0} & \mb{0}_{N_C\times N_{out}} \\
       P_n & D_{k:0}
     \end{bmatrix}
     \begin{bmatrix}
       \pmb{\pi}_A(\theta_p)_0 \\
       \pmb{\pi}_D(\theta_p)_0
     \end{bmatrix}
\end{equation}
Let us take the consensus among agents in $V_C$ as
\[
  \pmb{\pi}^*_A(\theta_p) 
    =(\mb{1}_{N_C}\mb{v}_A^T)\,\pmb{\pi}_A(\theta_p)_0.
\]
However, from \eqref{eq:pn_lim}, we know that $\lim_{n\to\infty} P_n=\mb{1}_{N_{out}}\mb{v}_A^T$. Then, by \eqref{eq:absorbing_chain2}, we can write 
\begin{align}
  \label{eq:absorbing_chain_consensus2}
  \begin{bmatrix}
    \pmb{\pi}^*_A(\theta_p) \\
    \pmb{\pi}^*_D(\theta_p)
  \end{bmatrix}
    =\underbrace{%
     \begin{bmatrix}
       \mb{1}_{N_C}\mb{v}_A^T 
         & \mb{0}_{N_C\times N_{out}} \\
       \mb{1}_{N_{out}}\mb{v}_A^T 
         & \mb{0}_{N_{out}\times N_{out}}
     \end{bmatrix}}_{%
     \mb{1}_{N_C+N_{out}}\mb{v}^{*^{T}}}
     \begin{bmatrix}
       \pmb{\pi}_A(\theta_p)_0 \\
       \pmb{\pi}_D(\theta_p)_0
     \end{bmatrix},
\end{align}
where $\mb{v}^*$ is a $(N_C+N_{out})$-sized stochastic vector created by concatenating the vectors $\mb{v}_A$ and $\mb{0}_{N_{out}\times 1}$, and 
\begin{equation}
  \label{eq:absorbing_chain_out_consensus}
  \pmb{\pi}_D^*(\theta_p) 
    =\mb{1}_{N_{out}} \mb{v}_A^T \pmb{\pi}_A(\theta_p)_0.
\end{equation}
Since \eqref{eq:absorbing_chain_consensus2} satisfies the conditions of Lemma~\ref{lem:consensus_evt}, all agents in $V$ achieve a consensus. Moreover, as \eqref{eq:absorbing_chain_consensus2} and \eqref{eq:absorbing_chain_out_consensus} show, this consensus is the same consensus achieved within $V_C$. 
\hspace*{\fill}\QEDclosed

An immediate consequence of Theorem~\ref{theorem:absorbing_chain} is 

\begin{mycoro}
\label{theorem:one_cautious_s_connected}
Consider the network $\mc{G}_k^{\dag}(\pmb{\varepsilon})=(V, E_k^{\dag}(\pmb{\varepsilon}))$ in \eqref{eq:Gdot} populated with receptively updating opinion followers and a single cautiously updating opinion leader. Suppose $D_k$ corresponds to the confidence matrix of the receptively updating opinion followers with $\Vert D_k\Vert\leq\rho<1,\,\forall k\in\mathbb{N}_{\geq 0}$. Then, with a CUE-based update strategy, all the agents reach a consensus opinion at the opinion of the opinion leader. 
\hspace*{\fill}\QEDopen
\end{mycoro}

\emph{Proof.}
Construct a 1-ODC as in Definition~\ref{def:1-ODC} with $\mc{G}_{C,k}^{\dag}(\pmb{\varepsilon})$ containing the opinion leader only and $\mc{G}_{out,k}^{\dag}(\pmb{\varepsilon})$ populated with the opinion followers. Proposition~\ref{prop:cautious} implies that $V_C^{\dag}$ (which consists of the only cautiously updating agent) generates a consensus. So, from Theorem~\ref{theorem:absorbing_chain}, all the agents in $V$ reach a consensus at the opinion of the opinion leader. 
\hspace*{\fill}\QEDclosed

\smallskip
\subsubsection{Two Opinion Leaders}
\label{subsec:2}

\begin{figure}[ht]
\centering
  \includegraphics[width=0.45\textwidth]{%
    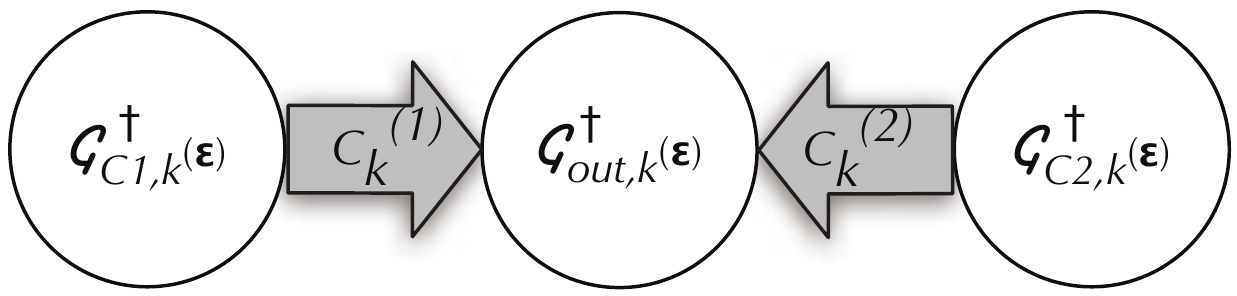}
  \caption{An opinion dynamics chain driven by two groups (2-ODC).}
  \label{fig:2C_chain}
\end{figure}

The situation turns out to be significantly more complicated when there are multiple opinion leaders. To look at the two opinion leader case, let us introduce 

\begin{mydef}[Opinion Dynamics Chain Driven By Two Groups (2-ODC)] 
\label{def:2-ODC}
The directed dynamic graph $\mc{G}_k^{\dag}(\pmb{\varepsilon})=(V, E_k^{\dag}(\pmb{\varepsilon}))$ in \eqref{eq:Gdot} is said to be an \emph{opinion dynamics chain driven by two groups (2-ODC)} if its corresponding confidence matrix $W_k$ in \eqref{eq:DTDsys} can be expressed as the lower block triangular matrix
\[
  W_k
    =\begin{bmatrix}
       A_k^{(1)} & \mb{0} & \mb{0} \\
       \mb{0} & A_k^{(2)} & \mb{0} \\
       C_k^{(1)} & C_k^{(2)} & D_k
     \end{bmatrix}. 
\]
Here, $A_k^{(1)}\in\mathbb{R}_{[0,1]}^{N_{C1}\times N_{C1}}$, $A_k^{(2)}\in\mathbb{R}_{[0,1]}^{N_{C2}\times N_{C2}}$, and $D_k\in\mathbb{R}_{[0,1]}^{N_{out}\times N_{out}}$; the other matrices have compatible sizes.
\hspace*{\fill}\QEDopen
\end{mydef}

One may view a 2-ODC as consisting of two `central' components $\mc{G}_{C1,k}^{\dag}(\pmb{\varepsilon})=(V_{C1}, E_{C1,k}^{\dag}(\pmb{\varepsilon}))$ (with $N_{C1}$ agents) and $\mc{G}_{C2,k}^{\dag}(\pmb{\varepsilon})=(V_{C2}, E_{C2,k}^{\dag}(\pmb{\varepsilon}))$ (with $N_{C2}$ agents), plus a third component $\mc{G}_{out,k}^{\dag}(\pmb{\varepsilon})=(V_{out}, E_{out,k}^{\dag}(\pmb{\varepsilon}))$ (with $N_{out}$ agents) s.t. no agent in $\mc{G}_{out,k}^{\dag}$ belongs to the in-components of either $\mc{G}_{C1,k}^{\dag}(\pmb{\varepsilon})$ or $\mc{G}_{C1,k}^{\dag}(\pmb{\varepsilon})$. See Fig.~\ref{fig:2C_chain}. Note that, $A_k^{(1)}$, $A_k^{(2)}$, and $D_k$ correspond to the confidence matrices of agents in $\mc{G}_{C1,k}^{\dag}(\pmb{\varepsilon})$, $\mc{G}_{C2,k}^{\dag}(\pmb{\varepsilon})$, and $\mc{G}_{out,k}^{\dag}(\pmb{\varepsilon})$, respectively. 

We now demonstrate that the agents in even a 2-ODC cannot reach a consensus in general, even though the agents in $V_{C1}$ and $V_{C2}$ may have achieved their own consensus opinions.

\begin{mythm}
\label{thm:2-ODC}
Consider agents embedded in a 2-ODC employing a CUE-based update strategy. Furthermore, suppose that 

  \tb{(a)}~$\lim_{n\to\infty} A_{n:0}^{(1)}=\mb{1}_{N_{C1}}\mb{v}_1^T$ and $\lim_{n\to\infty} A_{n:0}^{(2)}=\mb{1}_{N_{C2}}\mb{v}_2^T$, where $\mb{v}_1\in\mathbb{R}_{[0,1]}^{N_{C1}}$ and $\mb{v}_2\in\mathbb{R}_{[0,1]}^{N_{C2}}$ are stochastic vectors so that the agents in $V_{C1}$ and $V_{C2}$ achieve their own consensus opinions, and 
  
  \tb{(b)}~$\Vert D_k\Vert\leq\rho<1,\,\forall k\in\mathbb{N}_{\geq 0}$.
  
Then, the following are true:

  \tb{(i)}~The agents in $V$ (i.e., agents in $V_{C1}$, $V_{C2}$, and $V_{out}$) reach a consensus iff the consensus opinions of the agents in $V_{C1}$ and $V_{C2}$ are equal. 
  
  \tb{(ii)}~When the consensus opinions of the agents in $V_{C1}$ and $V_{C2}$ are not equal, a consensus among the agents in $V_{out}$ occurs if $\,\exists\,\lambda_k^{(1)}, \lambda_k^{(2)}\in (0,1)$ s.t. $\lambda_k^{(1)}+\lambda_k^{(2)}=1$ and $\lambda_k^{(1)}C_k^{(1)}\mb{1}_{N_{C1}}=\lambda_k^{(2)}C_k^{(2)}\mb{1}_{N_{C2}},\,\forall k\in\mathbb{N}_{\geq 0}$.
\hspace*{\fill}\QEDclosed
\end{mythm}

\emph{Proof.} 
Note that we may write
\begin{equation}
  \label{eq:WW}
  W_{n:0}
    =\begin{bmatrix}
       A_{n:0}^{(1)} & \mb{0} & \mb{0} \\
       \mb{0} & A_{n:0}^{(2)} & \mb{0} \\
       P_n^{(1)} & P_n^{(2)} & D_{n:0}
     \end{bmatrix},
\end{equation}
where, for $n\in\mathbb{N}_{\geq 0}$, 
\begin{align}
  \label{eq:rnplus1}
  P_{n+1}^{(1)}
    &=C_{n+1}^{(1)}A_{n:0}^{(1)}+D_{n+1}P_n^{(1)},\;
      P_0^{(1)}
        =C_0^{(1)}; 
      \notag \\
  P_{n+1}^{(2)}
    &=C_{n+1}^{(2)}A_{n:0}^{(2)}+D_{n+1}P_n^{(2)},\;
      P_0^{(2)}
        =C_0^{(2)}.
\end{align}

  \tb{(i)}~Suppose the two consensus opinions of the agents in $V_{C1}$ and $V_{C2}$ are equal. One may then recast the 2-ODC as a 1-DOC with the confidence matrices corresponding to the central component and the other component taken as 
\begin{math}
  \left[
    \begin{smallmatrix} 
      A_k^{(1)} & \mb{0} \\ 
      \mb{0} & A_k^{(2)} 
    \end{smallmatrix}
  \right]
\end{math} 
and $D_k$, respectively. Noting that the central component reaches a common consensus opinion, apply Theorem~\ref{theorem:absorbing_chain} to show that all the agents must reach a consensus which is identical to the common consensus opinion formed within the central component. Conversely, if the two consensus opinions of the agents in $V_{C1}$ and $V_{C2}$ are not equal, no consensus is possible among the agents in $V_{C1}$ and $V_{C2}$ because these two sets of agents do not update from each other. 

  \tb{(ii)}~Suppose the two consensus opinions of the agents in $V_{C1}$ and $V_{C2}$ are not equal. The row stochasticity of $W_k$ implies
\begin{alignat}{2}
  \label{eq:11}
  &\tb{1}_{N_{out}}
    &
      &=C_k^{(1)}\mb{1}_{N_{C1}}+C_k^{(2)}\mb{1}_{N_{C2}}
          +D_k\mb{1}_{N_{out}};
        \notag \\
  &\lambda_k^{(2)}\tb{1}_{N_{out}}  
    &
      &=C_k^{(1)}\mb{1}_{N_{C1}}+\lambda_k^{(2)}D_k\mb{1}_{N_{out}},\;
        k\in\mathbb{N}_{\geq 0},
\end{alignat}
where we used $\lambda_k^{(1)}C_k^{(1)}\mb{1}_{N_{C1}}=\lambda_k^{(2)}C_k^{(2)}\mb{1}_{N_{C2}}$ and $\lambda_k^{(1)} + \lambda_k^{(2)} = 1$. Now, proceed as we did in the proof of Theorem~\ref{theorem:absorbing_chain}. Subtract $\lambda_k^{(2)}\mb{1}_{N_{out}}\mb{v}_1^T$ from both sides of \eqref{eq:rnplus1} and substitute for $\mb{1}_{N_{out}}$ from \eqref{eq:11}: 
\begin{multline*}
  \hspace*{-0.1in}
  P_{n+1}^{(1)}-\lambda_k^{(2)}\mb{1}_{N_{out}}\mb{v}_1^T \\
    =C_{n+1}^{(1)}[A_{n:0}^{(1)}-\mb{1}_{N_{C1}}\mb{v}_1^T]  
       +D_{n+1}
        [P_n^{(1)}-\lambda_k^{(2)}\mb{1}_{N_{out}}\mb{v}_1^T],\hspace*{-0.1in}
\end{multline*}
for $n\in\mathbb{N}_{\geq 0}$. As before, we use the notation $\Delta P_n^{(1)}=P_n^{(1)}-\lambda_k^{(2)}\mb{1}_{N_{out}}\mb{v}_1^T$ to express this as 
\[
  \Delta P_{n+1}^{(1)}
    =C_{n+1}^{(1)}[A_{n:0}^{(1)}-\mb{1}_{N_{C1}}\mb{v}_1^T]
       +D_{n+1}\Delta P_n^{(1)},      
\]
for $n\in\mathbb{N}_{\geq 0}$. Now, bound $\Vert\Delta P_{n+1}^{(1)}-D_{n+1}\Delta P_n^{(1)}\Vert$ as
\begin{align*}
  &\vert 
     \Vert\Delta P_{n+1}^{(1)}\Vert
       -\Vert D_{n+1}\Delta P_n^{(1)}\Vert 
   \vert
   \notag \\
  &\quad
     \leq
      \Vert\Delta P_{n+1}^{(1)}-D_{n+1}\Delta P_n^{(1)}\Vert 
     =\Vert 
        C_{n+1}^{(1)}[A_{n:0}^{(1)}-\mb{1}_{N_{C1}}\mb{v}_1^T]
      \Vert 
   \notag \\
  &\quad
     \leq 
      \Vert A_{n:0}^{(1)}-\mb{1}_{N_{C1}}\mb{v}_1^T \Vert,
\end{align*}
where we used the sub-stochasticity of $C_n^{(1)}$. 

Now, as in the proof of Theorem~\ref{theorem:absorbing_chain}, use the fact that the agents in $V_{C1}$ and $V_{C2}$ each achieve a consensus to show that
\begin{align*}
  \lim_{n\to\infty} 
  \Vert P_n^{(1)}-\lambda_k^{(2)}\mb{1}_{N_{out}}\mb{v}_1^T\Vert
    &=\lim_{n\to\infty} 
      \Vert P_n^{(2)}-\lambda_k^{(1)}\mb{1}_{N_{out}}\mb{v}_2^T\Vert
      \notag \\
    &=0.
\end{align*}
In other words, 
\[
  \lim_{n\to\infty} P_n^{(1)}
    =\lambda_k^{(2)} \mb{1}_{N_{out}}\mb{v}_1^T;\;\;
  \lim_{n\to\infty} P_n^{(2)}
    =\lambda_k^{(1)} \mb{1}_{N_{out}} \mb{v}_2^T. 
\]

Denote the consensus among the agents in $V_{C1}$ and $V_{C2}$ as
\begin{align*}
  \pmb{\pi}^*_{A^{(1)}}(\theta_p)
    &=(\mb{1}_{N_{C1}}\mb{v}_1^T)\,
      \pmb{\pi}_{A^{(1)}}(\theta_p)_0;
      \notag \\
  \pmb{\pi}^*_{A^{(2)}}(\theta_p)
    &=(\mb{1}_{N_{C2}}\mb{v}_2^T)\,
      \pmb{\pi}_{A^{(2)}}(\theta_p)_0, 
\end{align*}
respectively. Then we have
\[
  \arraycolsep=1.8pt\def\arraystretch{1.0}
  \begin{bmatrix}
    \pmb{\pi}^*_{A^{(1)}}(\theta_p) \\
    \pmb{\pi}^*_{A^{(2)}}(\theta_p) \\
    \pmb{\pi}^*_D(\theta_p)
  \end{bmatrix}\!
    =\!
     \begin{bmatrix}
       \mb{1}_{N_{C1}}\mb{v}_1^T & \mb{0} & \mb{0} \\
       \mb{0} & \mb{1}_{N_{C2}}\mb{v}_2^T & \mb{0} \\
       \lambda_k^{(2)}\mb{1}_{N_{out}}\mb{v}_1^T 
         & \lambda_k^{(1)}\mb{1}_{N_{out}}\mb{v}_2^T & \mb{0}
     \end{bmatrix}\!\!
     \begin{bmatrix}
       \pmb{\pi}_{A^{(1)}}(\theta_p)_0 \\
       \pmb{\pi}_{A^{(2)}}(\theta_p)_0 \\
       \pmb{\pi}_D(\theta_p)_0
     \end{bmatrix}\!,
\]
where $\pmb{\pi}^*_D(\theta_p)$ is given as
\[
  \pmb{\pi}^*_D(\theta_p)
    =\mb{1}_{N_{out}}
     \begin{bmatrix}
       \lambda_k^{(2)}\mb{v}_1^T & \lambda_k^{(1)}\mb{v}_2^T
     \end{bmatrix}
     \begin{bmatrix}
       \pmb{\pi}_{A^{(1)}}(\theta_p)_0 \\
       \pmb{\pi}_{A^{(2)}}(\theta_p)_0
     \end{bmatrix}.
\]
Hence, from Lemma~\ref{lem:consensus_evt}, we conclude that the agents in $V_{out}$ reach a consensus if $\lambda_k^{(1)}+\lambda_k^{(2)}=1,\,\forall k\in\mathbb{N}_{\geq 0}$. 
\hspace*{\fill}\QEDclosed

One may interpret this result in the following manner: the matrices $C_k^{(1)}$ and $C_k^{(2)}$ signify the `weights' or `bias' that agents in $V_{out}$ give to the agents in $V_{C1}$ and $V_{C2}$, respectively. To reach a consensus, each agent in $V_{out}$ must give the same proportion of weights to the agents in $V_{C1}$ and $V_{C2}$ for all $k\in\mathbb{N}_{\geq 0}$: $\lambda_k^{(2)}>\lambda_k^{(1)}$ implies that a higher weight is given to the opinions of the agents in $V_{C1}$ than to the opinions of the agents in $V_{C2}$. This might be due to stronger interconnections between the agents in $V_{C1}$ and $V_{out}$ or it could simply be due to a bias towards the opinions of the agents in $V_{C1}$. 

As an immediate consequence of Theorem~\ref{thm:2-ODC} we get

\begin{mycoro}
\label{theorem:two_cautious_s_connected}
Consider the network $\mc{G}_k^{\dag}(\pmb{\varepsilon})=(V, E_k^{\dag}(\pmb{\varepsilon}))$ in \eqref{eq:Gdot} populated with receptively updating opinion followers and two cautiously updating opinion leaders. Suppose $D_k$ corresponds to the confidence matrix of the receptively updating opinion followers with $\Vert D_k\Vert\leq\rho<1,\,\forall k\in\mathbb{N}_{\geq 0}$. Then, with a CUE-based update strategy, the following are true:

  \tb{(i)}~All the agents reach a consensus iff the opinions of the two opinion leaders are equal. 
  
  \tb{(ii)}~When the opinions of the opinion leaders are not equal, the opinion followers form an opinion cluster if the proportion of weights each receptively updating agent gives to the two opinion leaders is identical and no opinion leader is given zero weight. 
\hspace*{\fill}\QEDclosed
\end{mycoro}


\subsection{Dirichlet Agent Opinions}
\label{subsec:DirichletBBA}


Next, let us suppose that agents opinions are modeled via Dirichlet BoEs. Then, using the properties $Bl(\theta_i|\theta_i)=1$, $Bl(\theta_i|\theta_j)=0,\,i\neq j$, and $Bl(B|\Theta)=B,\,\forall B\subseteq\Theta$, one can easily show that the CUE-based update mechanism in Definition~\ref{def:CUEm} retains the Dirichlet property of the updated BoEs at each step. For this Dirichlet BoE case, the CUE-based opinion update reduces to the following DT dynamic system:
\begin{equation}
  \label{eq:DTDsysDirichlet}
  \pmb{\pi}(\theta_p)_{k+1} 
    =\breve{W}_k\,\pmb{\pi}(\theta_p)_k,\;
     p\in\ol{1,M}.
\end{equation}
Here $\pmb{\pi}(\cdot)_k$ is as in \eqref{eq:DTDsys} and $\breve{W}_k=\{\breve{w}_{ij,k}\}\in\mathbb{R}^{N\times N}$ where $\breve{w}_{ij,k},\, i,j\in\ol{1,N},\,k\in\mathbb{N}_{\geq 0}$. When the $i$-th agent is receptively updating,  
\begin{equation}
  \label{eq:wijk_breve_receptive}
  \breve{w}_{ij,k} 
    =\begin{cases}
       \alpha_{i,k}, 
         & \text{for $i=j$}; \\
       \displaystyle
       \frac{(1-\alpha_{i,k})(1+m_j(\Theta)_k)}
            {\vert\mc{N}_{i,k}(\varepsilon_i)\vert}, 
         & \text{for $j\in\mc{N}_{i,k}(\varepsilon_i)$}; \\
       0, 
         & \text{otherwise};
     \end{cases}
\end{equation} 
when the $i$-th agent is cautiously updating, 
\begin{equation}
  \label{eq:wijk_breve_cautious}
  \breve{w}_{ij,k} 
    =\begin{cases}
       1, 
         & \text{for $i=j$};\\
       \displaystyle
       \frac{(1-\alpha_{i,k})m_i(\Theta)_k}
            {\vert\mc{N}_{i,k}(\varepsilon_i)\vert}, 
         & \text{for $j\in\mc{N}_{i,k}(\varepsilon_i)$}; \\
       0, 
         & \text{otherwise}.
     \end{cases}
\end{equation}
While we may still refer to $\breve{W}_k$ as the corresponding confidence matrix, unlike $W_k$, $\breve{W}_k$ is not necessarily stochastic. 

To proceed, we take inspiration from \cite{Pre09_ICIF}, where it is shown that, under mild conditions, the masses for complete ambiguity $\Theta$ vanish when two Dirichlet agents mutually update each other. The same result turns out to hold true for multiple BoEs.

\begin{mylem}
\label{lem:Dirichlet}
Consider the CUE-based updating of the Dirichlet BoEs as in \eqref{eq:DTDsysDirichlet}. If 
\[
  \alpha_{i,k}+\sum_{j\neq i} \beta_{ij}(\Theta)_k
    \leq\rho
    <1,\,
     \forall i,j\in \ol{1,N},\,
     \forall k\in\mathbb{N}_{\geq 0}, 
\]
then $\lim_{n\to\infty} m_i(\Theta)_n=0,\,\forall i\in\ol{1,N}$.
\hspace*{\fill}\QEDopen
\end{mylem}

\emph{Proof.}
Observe that the update of $\pmb{\pi}(\Theta)$ can be written as
\[
  \pmb{\pi}(\Theta)_{k+1} 
    =\pmb{\Gamma}_k \pmb{\pi}(\Theta)_k,
\]
where 
\begin{align*}
  \pmb{\Gamma}_k 
    =\begin{bmatrix}
	   \alpha_{1,k} & \beta_{12}(\Theta)_k 
	     & \beta_{13}(\Theta)_k & \cdots & \beta_{1N}(\Theta)_k \\
	   \beta_{21}(\Theta)_k	& \alpha_{2,k} 
	     & \beta_{23}(\Theta)_k	& \cdots & \beta_{2N}(\Theta)_k \\
	   \vdots & \vdots & \ddots & \vdots & \vdots \\	
	   \beta_{N1}(\Theta)_k	& \beta_{N2}(\Theta)_k 
	     & \beta_{N3}(\Theta)_k & \cdots & \alpha_{N,k} 
	 \end{bmatrix}.
\end{align*}
The condition in the statement implies that $\Vert\pmb{\Gamma}_k\Vert_{\infty}\leq\rho<1,\,\forall k\in\mathbb{N}_{\geq 0}$, which guarantees the claim.
\hspace*{\fill}\QEDclosed

Note that, each BoE being updated possessing at least one singleton focal element ensures that the condition in Lemma~\ref{lem:Dirichlet} is satisfied, which in turn, ensures that the mass for each completely ambiguous proposition vanishes in the limit. When this occurs, one may show the following results which apply to Dirichlet agents. These are the counterparts to Corollaries~\ref{theorem:one_cautious_s_connected} and \ref{theorem:two_cautious_s_connected} which apply to probabilistic agents. Rigorous proofs of these results are not provided; they can be carried out in a manner which parallel the proofs of Corollaries~\ref{theorem:one_cautious_s_connected} and \ref{theorem:two_cautious_s_connected}.

\begin{mycoro}
\label{theorem:one_cautious_s_connected_Dirichlet}
Consider the network $\mc{G}_k^{\dag}(\pmb{\varepsilon})=(V, E_k^{\dag}(\pmb{\varepsilon}))$ in \eqref{eq:Gdot} populated with receptively updating Dirichlet opinion followers and a single cautiously updating Dirichlet opinion leader. Suppose that the condition in Lemma~\ref{lem:Dirichlet} is satisfied. Let $D_k$ denote the confidence matrix of the receptively updating opinion followers with $\Vert D_k\Vert\leq\rho<1,\,\forall k\geq N_D$, for some $N_D\in\mathbb{N}_{\geq 0}$. Then all the agents reach a consensus opinion. 
\hspace*{\fill}\QEDclosed
\end{mycoro}

\begin{mycoro}
\label{theorem:two_cautious_s_connected_Dirichlet}
Consider the network $\mc{G}_k^{\dag}(\pmb{\varepsilon})=(V, E_k^{\dag}(\pmb{\varepsilon}))$ in \eqref{eq:Gdot} populated with receptively updating Dirichlet opinion followers and two cautiously updating Dirichlet opinion leaders. Suppose that the condition in Lemma~\ref{lem:Dirichlet} is satisfied. Let $D_k$ denote the confidence matrix of the receptively updating opinion followers with $\Vert D_k\Vert\leq\rho<1,\,\forall, k\geq N_D$, for some $N_D\in\mathbb{N}_{\geq 0}$. Then, with a CUE-based update strategy, the following are true:

  \tb{(i)}~All the agents reach a consensus iff the converged opinions of the two opinion leaders are equal. 
  
  \tb{(ii)}~When the opinions of the opinion leaders are not equal, the opinion followers form an opinion cluster if the proportion of weights each receptively updating agent gives to the two opinion leaders is identical for $k\geq N_D$ and no opinion leader is given zero weight. 
\hspace*{\fill}\QEDclosed
\end{mycoro}


\section{Empirical Evaluations and Discussion}
\label{sec:Results} 


We now present some typical results obtained through extensive simulations of scenarios where agent opinions are captured via p.m.f.s (see Section~\ref{PAOs}), Dirichlet BoEs (see Section~\ref{DAOs}), and more general DST BoEs (see Section~\ref{DSTAOs}). The results confirm our theoretical analysis in Section~\ref{sec:ODC} and demonstrate the applicability of these results for the more general DST models presented in Section~\ref{sec:ourModel}. 

The agents in all the simulations employ a CUE-based opinion update strategy with $\alpha_i=0.50,\,\forall i\in\ol{1,N}$. The agents' bounds of confidence are taken as identical, i.e., $\varepsilon_i=\varepsilon,\,\forall i\in\ol{1,N}$. Note that, for $\alpha_i>0,\,\forall i\in\ol{1,N}$, and for sufficiently large $\varepsilon$, all agents satisfy the self-communicating or the strong-aperiodic property \cite{Lorenz2006, Touri_2014}. Even though being embedded in a static network, the agents must accommodate the bounds of confidence as the opinions are updated. In effect, this creates a dynamic network $\mc{G}_k^{\dag}(\pmb{\varepsilon})$.

For ease of visualization, consensus/cluster formation are displayed using bifurcation diagrams that depict the state of consensus/cluster formation in the limit density versus $\varepsilon$ \cite{Lorenz2007}. 


\subsection{Probabilistic Agent Opinions} 
\label{PAOs}


For our simulations we embed seven agents on a graph of seven nodes and  100 agents on  Erd\H{o}s-R\'{e}nyi (ER) random graphs  of 100 nodes. 

\subsubsection{Simulations with Seven Agents}
\label{sec:SimulationsWithSevenNodes}

\begin{figure}[htpb]
  \centering 
  \subfigure[Network topology.\label{fig:A_simu_nx}]{%
    \includegraphics[width=0.18\textwidth]{%
      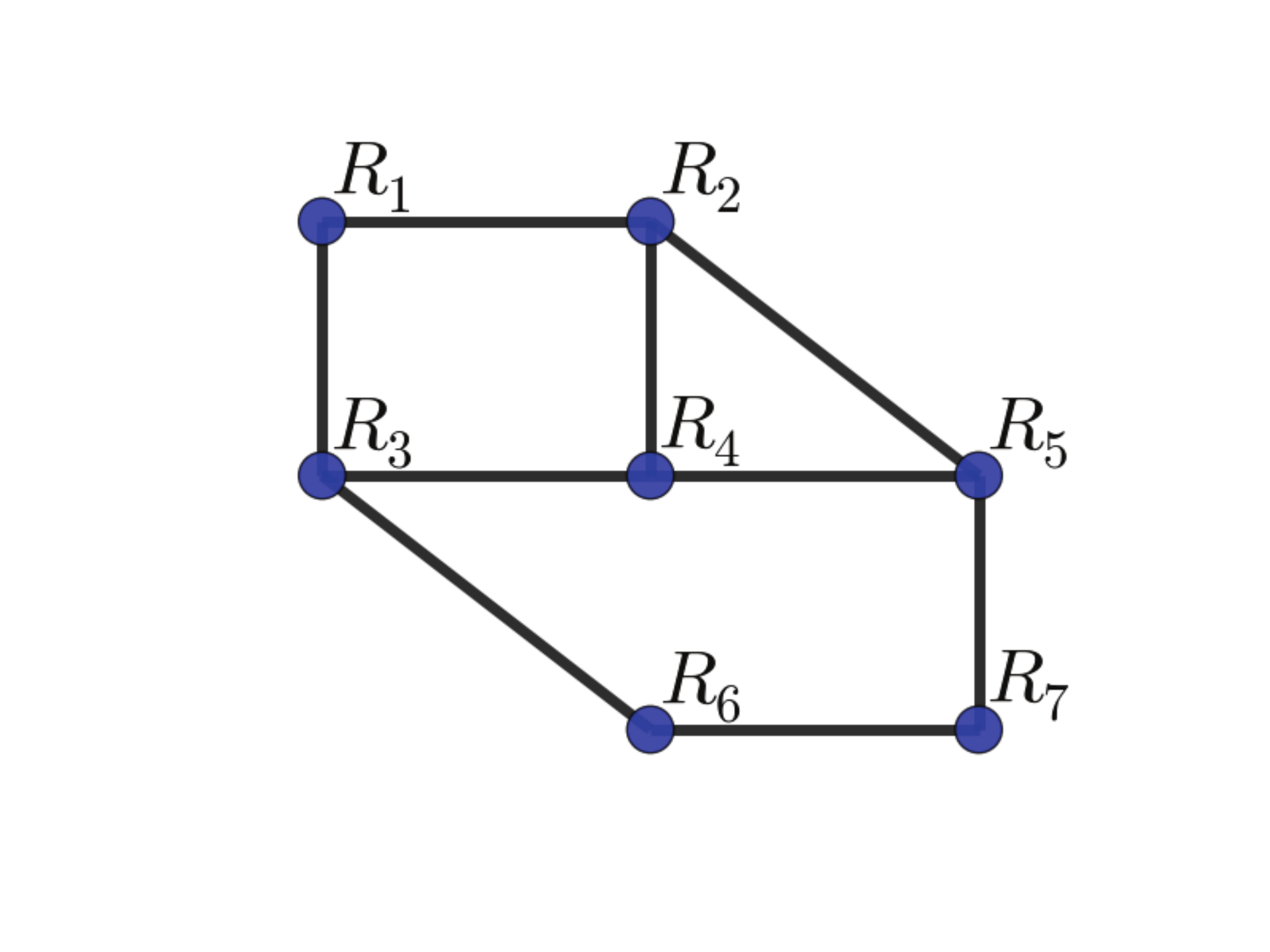}} 
  \subfigure[Bifurcation diagram for $\pmb{\pi}(\theta_1)$.\label{fig:A_simu_bifur}]{%
    \includegraphics[width=0.30\textwidth]{%
      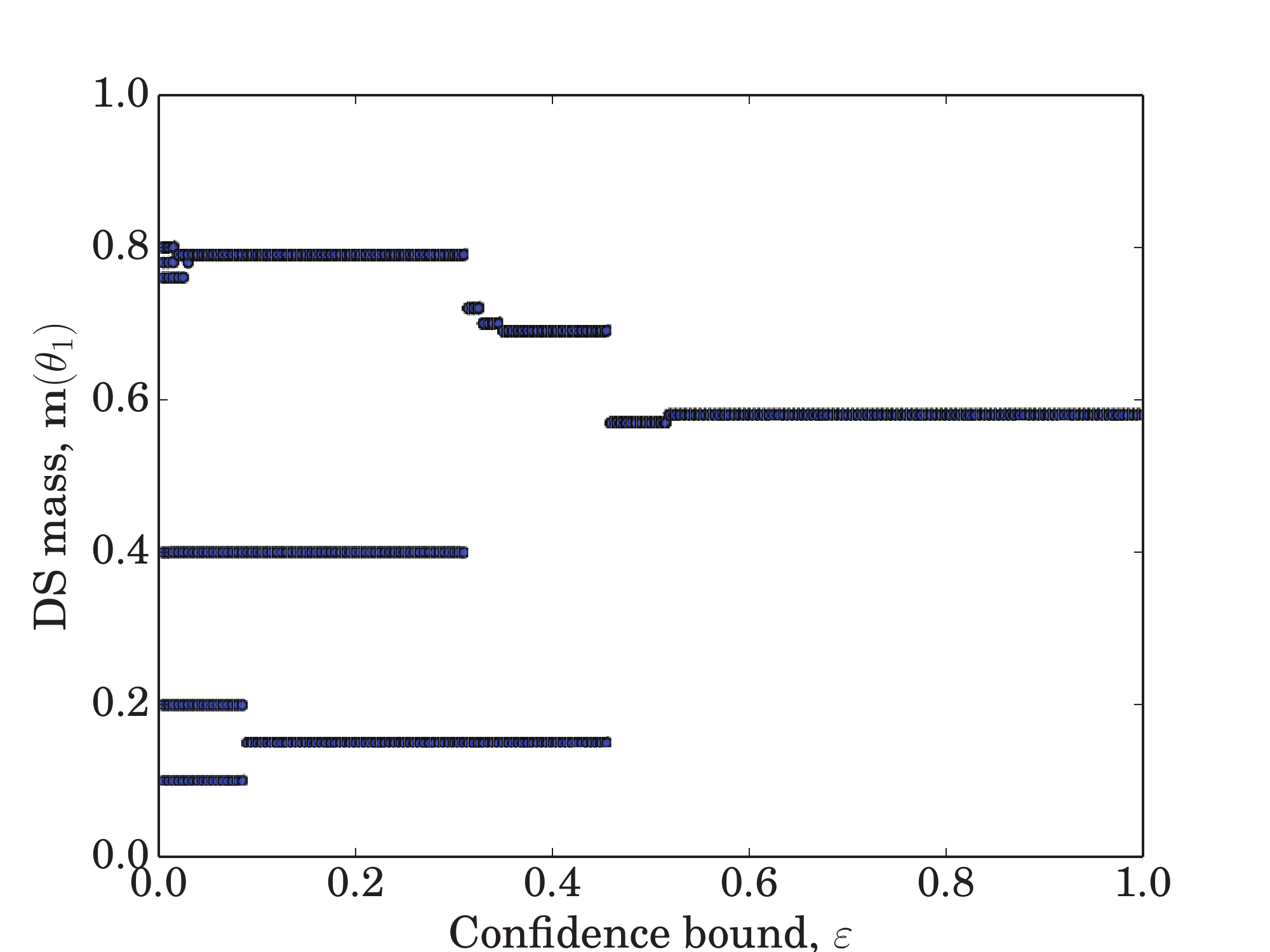}}
  \subfigure[For $\varepsilon=0.5$, first iteration: $(R_3,\, R_6)$ and $(R_5,\, R_7)$ do not exchange opinions because the corresponding opinion distances are above $\varepsilon=0.5$.\label{fig:A_iteration_0}]{%
    \includegraphics[width=0.23\textwidth]{%
      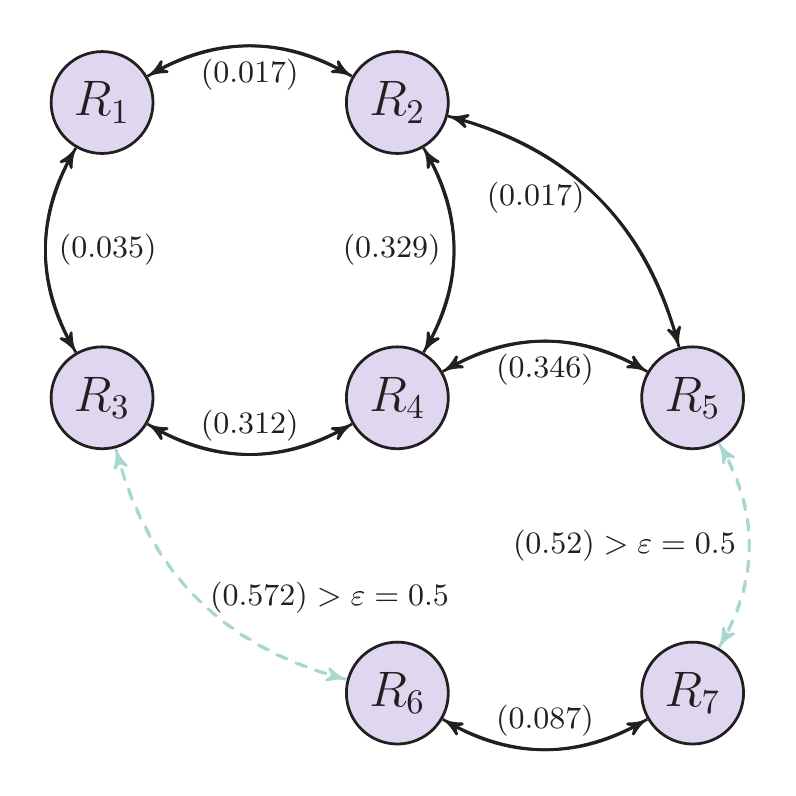}}\; 
  \subfigure[For $\varepsilon=0.5$, second iteration: All neighbor agents exchange opinions because all corresponding opinion distances are below $\varepsilon=0.5$.\label{fig:A_iteration_1}]{%
    \includegraphics[width=0.23\textwidth]{%
      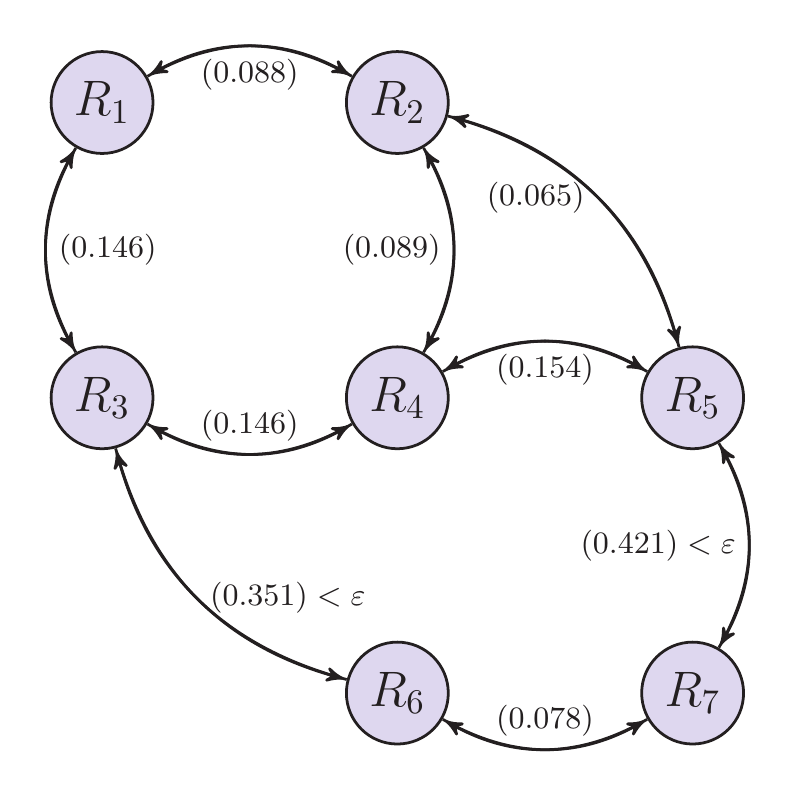}}     
  \caption{Probabilistic agents: Simulation results for seven receptively updating agents ($R_i,\,i\in\ol{1,7}$) and no opinion leaders. A consensus occurs for $\varepsilon>0.46$ (approx.). For $\varepsilon=0.5$, Figs~\ref{fig:A_iteration_0} and \ref{fig:A_iteration_1} show the `directions' of opinion exchange at the first two iterations. Edge labels indicate the distance between the opinions of the corresponding agent pair.} 
  \label{fig:A_simu}
\end{figure}

\begin{figure}[htpb]
  \centering 
  \subfigure[Network topology.\label{fig:C_simu_nx}]{%
    \includegraphics[width=0.18\textwidth]{%
      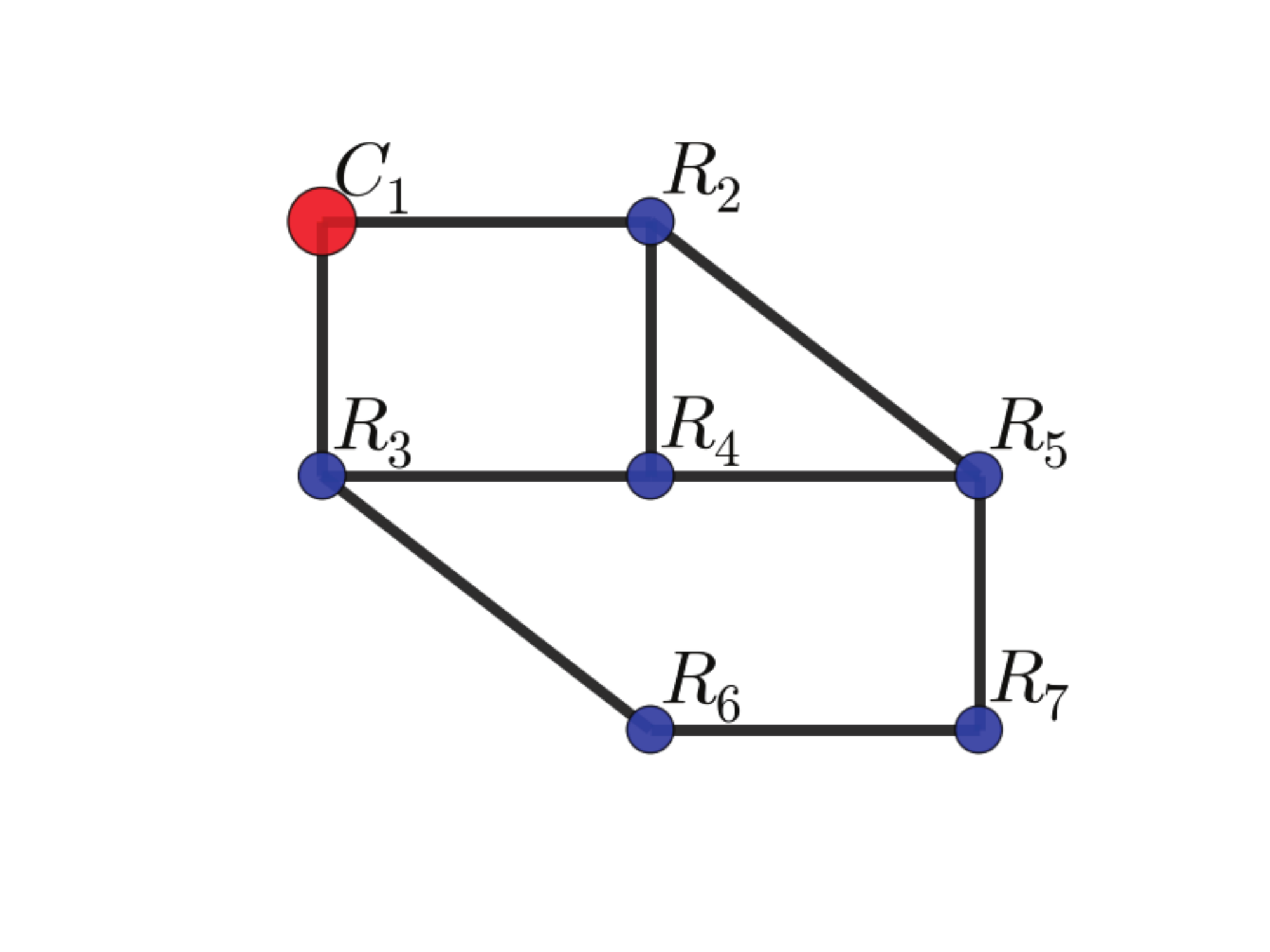}}
  \subfigure[Bifurcation diagram for $\pmb{\pi}(\theta_1)$.\label{fig:C_simu_bifur}]{%
    \includegraphics[width=0.3\textwidth]{%
      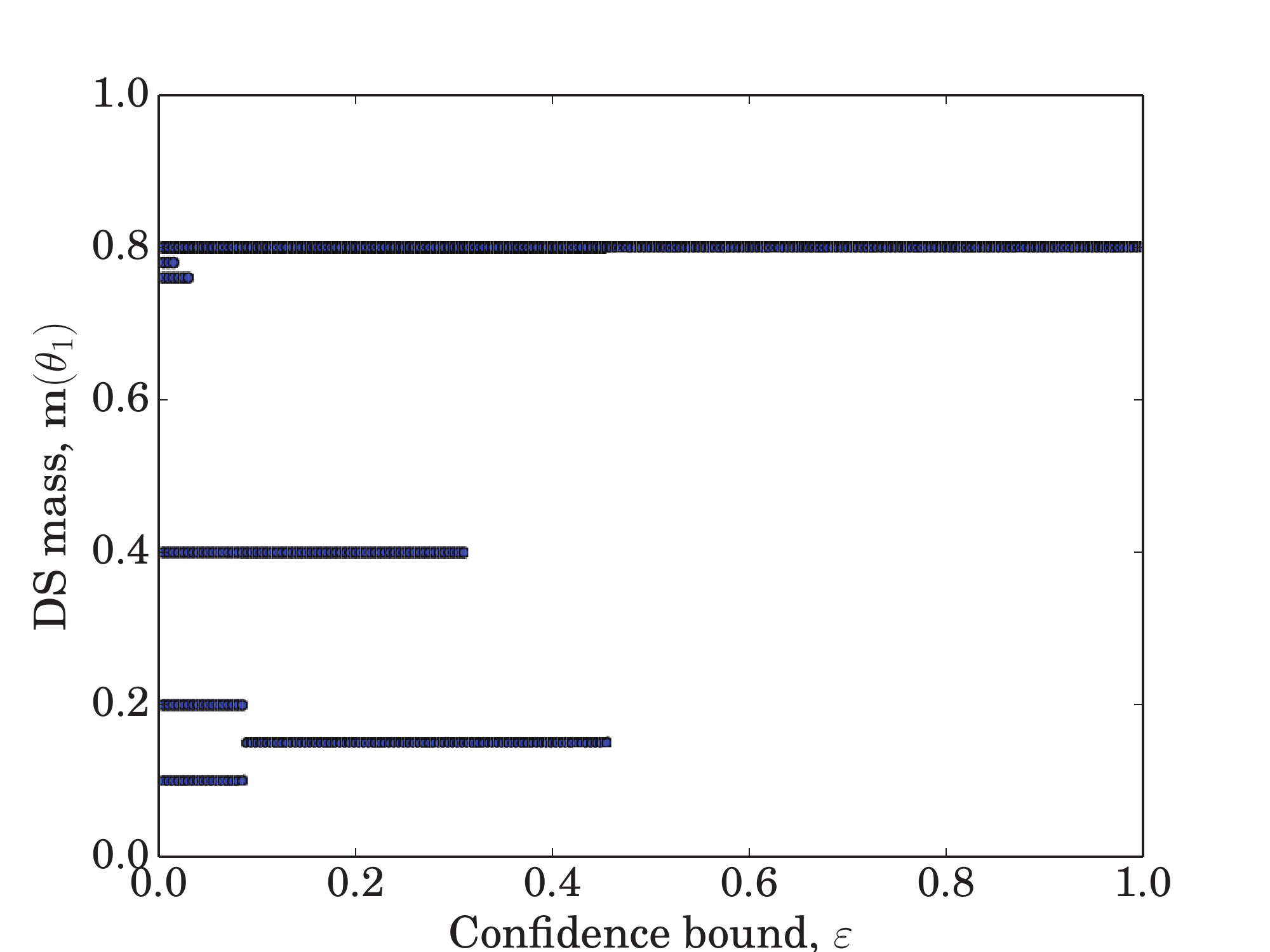}}	
  \subfigure[For $\varepsilon=0.46$, first iteration: $(R_3,\, R_6)$ and $(R_5,\, R_7)$ do not exchange opinions.\label{fig:C_iteration_0}]{%
    \includegraphics[width=0.15\textwidth]{%
      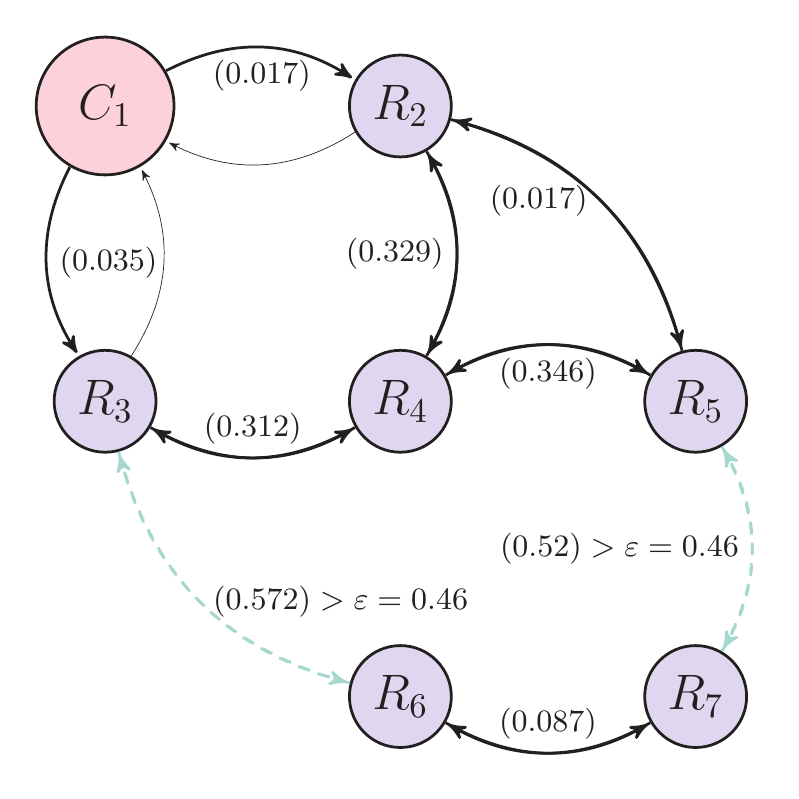}}\; 
  \subfigure[For $\varepsilon=0.46$, second iteration. $(R_3,\, R_6)$ now exchange opinions; $(R_5,\, R_7)$ still do not exchange opinions.\label{fig:C_iteration_1}]{%
    \includegraphics[width=0.15\textwidth]{%
      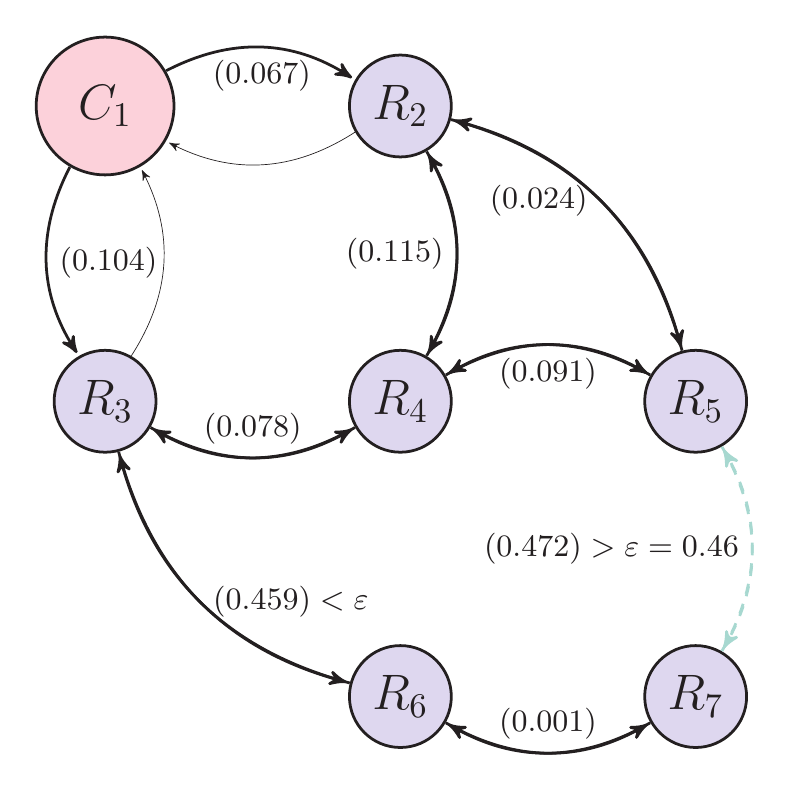}}\;     
  \subfigure[For $\varepsilon=0.46$, third iteration. All agents exchange opinions with their neighbors.\label{fig:C_iteration_2}]{%
    \includegraphics[width=0.15\textwidth]{%
      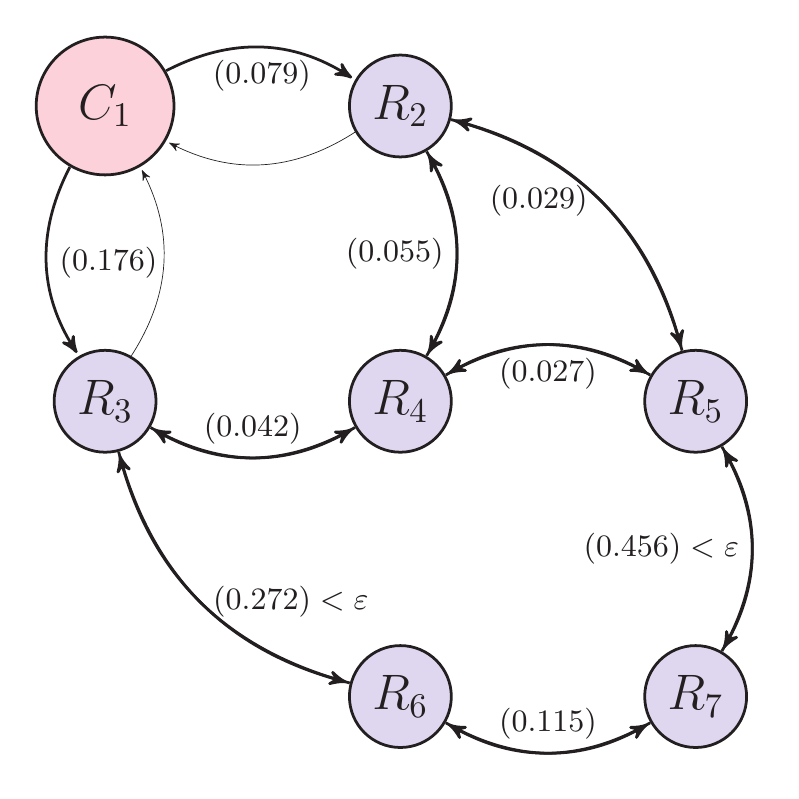}}  
  \caption{Probabilistic agents: Simulation results for one opinion leader ($C_1$) and six receptively updating agents ($R_i,\,i\in\ol{2,7}$). A consensus occurs for $\varepsilon>0.46$ (approx.), at $C_1$'s opinion (i.e., $m(\theta_1)=0.80$). For $\varepsilon=0.46$, Figs~\ref{fig:C_iteration_0}, \ref{fig:C_iteration_1}, and \ref{fig:C_iteration_2} show the `directions' of opinion exchange at the first three iterations. Edge labels indicate the distance between the opinions of the corresponding agent pair.} 
  \label{fig:C_simu}
\end{figure}

\begin{figure}[htpb]
  \centering 
  \subfigure[Network topology.\label{fig:CC_nx_ori}]{%
    \includegraphics[width=0.18\textwidth]{%
      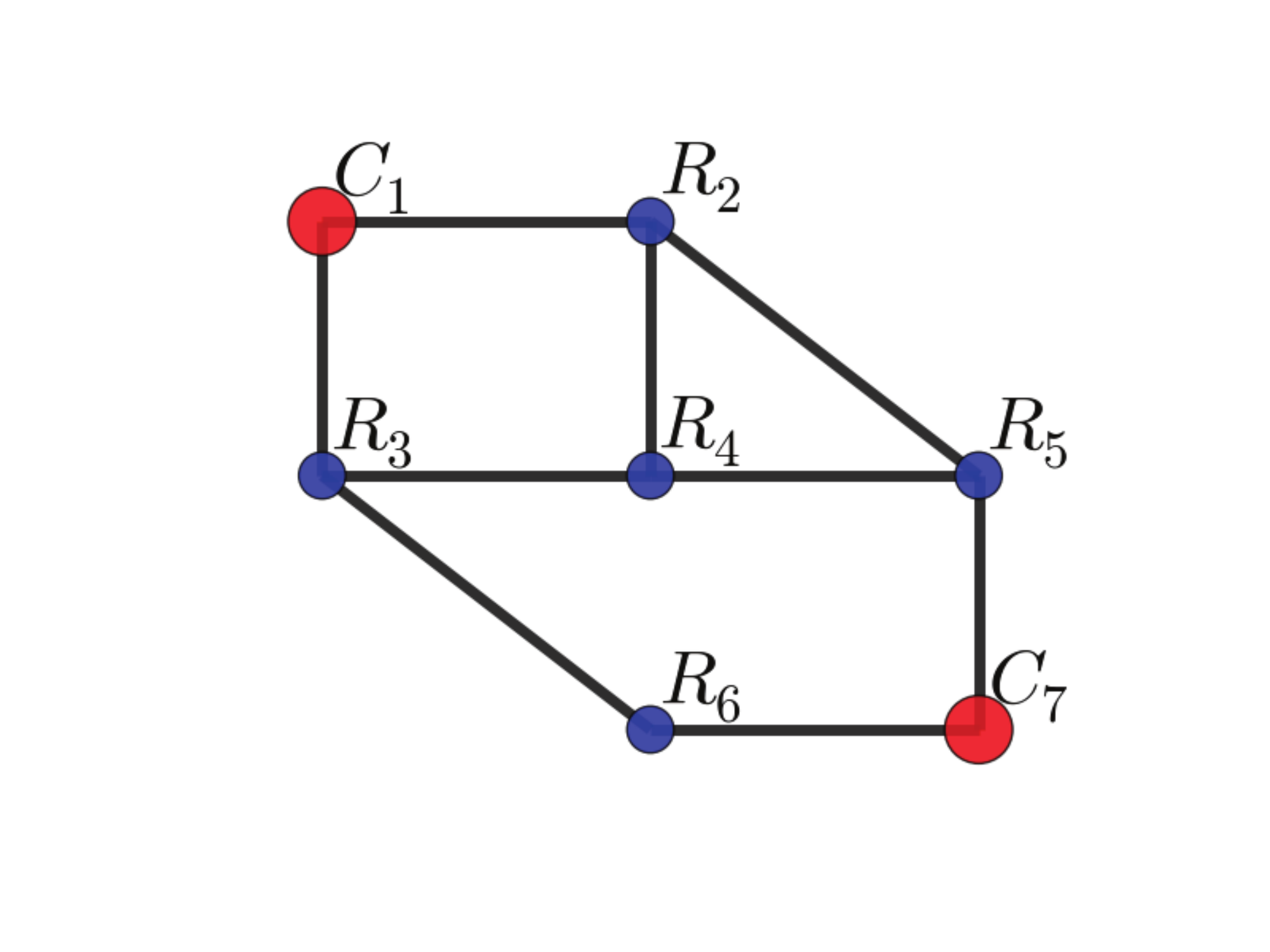}}
  \subfigure[Bifurcation diagram for $\pmb{\pi}(\theta_1)$.\label{fig:CC_simu_bifur}]{%
    \includegraphics[width=0.3\textwidth]{%
      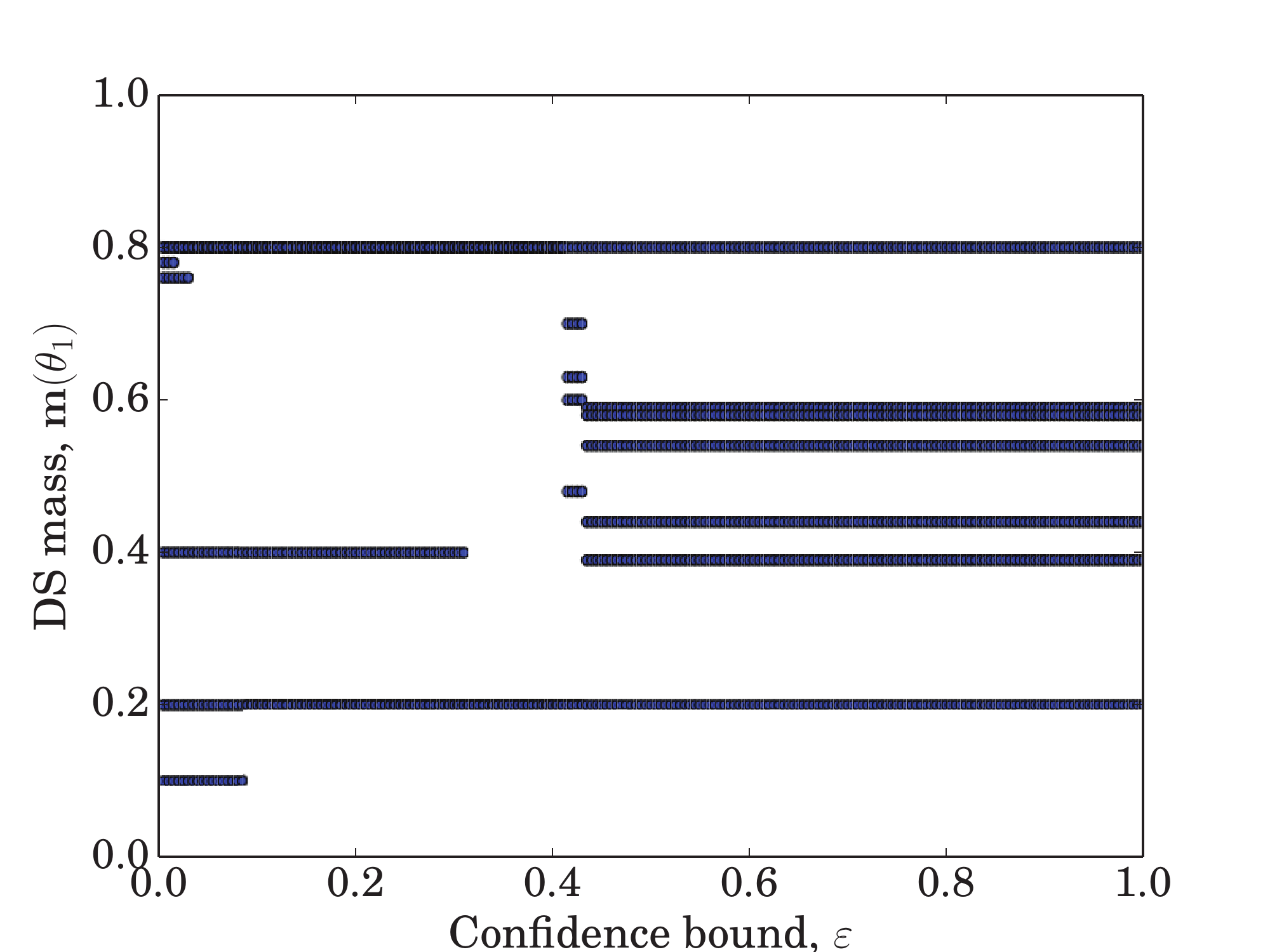}}
  \caption{Probabilistic agents: Simulation results for  two opinion leaders ($C_1$ and $C_7$) and five receptively updating agents ($R_i,\,i\in\ol{2,6}$). With the two opinion leaders possessing different opinions, no consensus is achieved. A minimum of $2$ opinion clusters are achieved for $0.31<\varepsilon <0.42$ (approx.). There is no consensus among the receptively updating agents.} 
  \label{fig:CC_simu}
\end{figure}

\begin{figure}[htpb]
  \centering 
   \subfigure[Altered topology.\label{fig:CC_nx_alt}]{%
     \includegraphics[width=0.18\textwidth]{%
       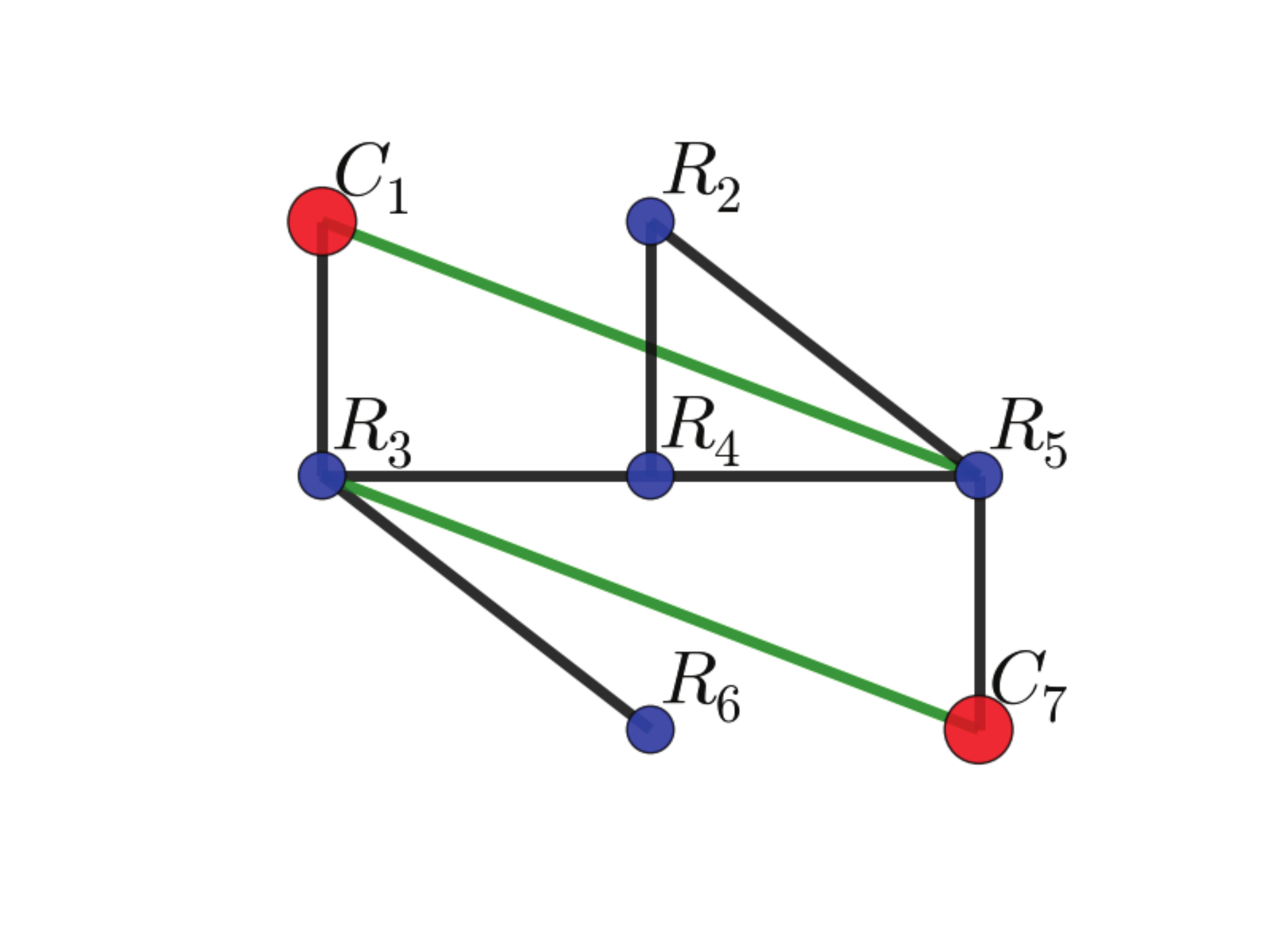}}
   \subfigure[Bifurcation diagram for $\pmb{\pi}(\theta_1)$.\label{fig:CC_simu_bifur_alt}]{%
   \includegraphics[width=0.3\textwidth]{%
     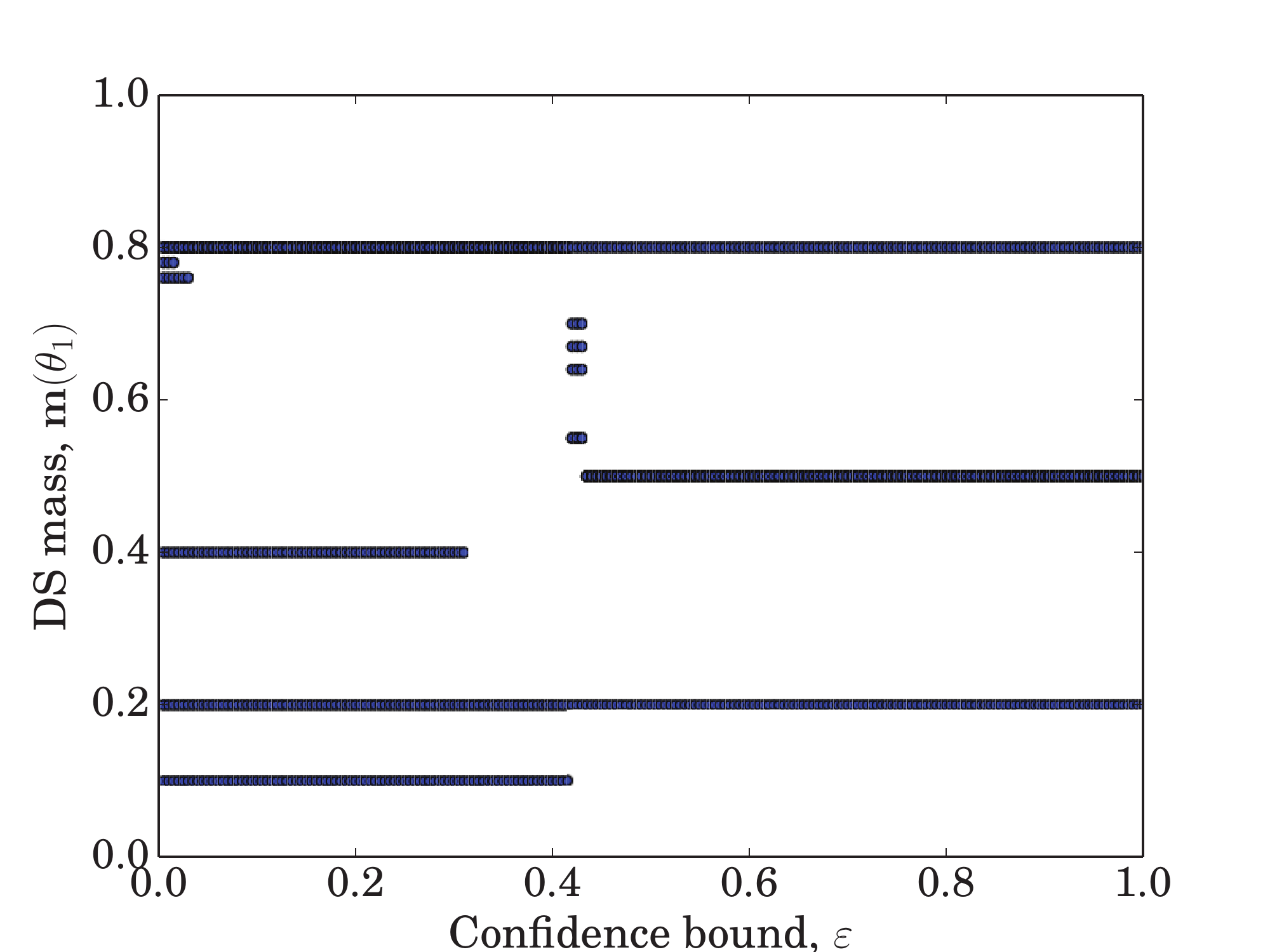}}
   \caption{Probabilistic agents: Simulation results for  two opinion leaders ($C_1$ and $C_7$) and five receptively updating agents ($R_i,\,i\in\ol{2,6}$) embedded in a topology that generates a consensus among the receptively updating agents. This consensus appears for $\varepsilon> 0.43$ (approx.).} 
  \label{fig:CC_simu_altered}
\end{figure}

The FoD of the opinion BoE of each agent is $\Theta=\{\theta_1, \theta_2, \theta_3\}$. For the results shown in Figures~\ref{fig:A_simu}~to~\ref{fig:CC_simu_altered}, initial opinion profile of $\theta_1$ is selected as $\pmb{\pi}(\theta_1)_0=[0.80, 0.78, 0.76, 0.40, 0.80, 0.10, 0.20]^T$; the remaining masses are equally distributed among the opinion profiles $\pmb{\pi}(\theta_2)_0$ and $\pmb{\pi}(\theta_3)_0$. 

\textbf{\emph{No Opinion Leaders.}} 
Fig.~\ref{fig:A_simu_nx} shows the network topology of the seven receptively updating agents $R_i,\,i\in\ol{1,7}$. As the corresponding bifurcation diagram in Fig.~\ref{fig:A_simu_bifur} shows, for smaller values of $\varepsilon$, seven `opinion clusters' are generated because agents are essentially isolated. 
As the value of  $\varepsilon$ is increased, the number of opinion clusters decreases because the agents  are updating their opinions based on opinions of their neighbors who are within their confidence bounds. Eventually, for $\varepsilon>0.46$ (approx.), as expected from our analytical results in Section~\ref{subsec:0}, a consensus emerges.

Even though the underlying network topology shown in Fig.~\ref{fig:A_simu_nx} is static, opinion updating occurs in a dynamic network $\mc{G}_k^{\dag}(\pmb{\varepsilon})$. To illustrate this further, consider the case when $\varepsilon=0.5$. As Fig.~\ref{fig:A_iteration_0} illustrates, agent pairs $(R_3,\, R_6)$ and $(R_5,\, R_7)$ do not exchange opinions because their opinion distances exceed the bound of confidence $\varepsilon=0.5$. However, after the first iteration of opinion exchanges, distances among agents change. As Fig.~\ref{fig:A_iteration_1} illustrates, the opinion distances of all agents (including $(R_3,\, R_6)$ and $(R_5,\, R_7)$) are now well within $\varepsilon=0.5$ and, at the second iteration, all agents exchange opinions with their neighboring agents.

\textbf{\emph{Single Opinion Leader.}} 
We create this scenario by replacing agent $R_1$ in Fig.~\ref{fig:A_simu_nx} with a cautiously updating agent $C_1$ thus obtaining the graph in Fig.~\ref{fig:C_simu_nx}. For this topology the bifurcation diagram for $\pmb{\pi}(\theta_1)$ is depicted in Fig.\ref{fig:C_simu_bifur}. As before, for smaller values of $\varepsilon$, each agent forms its own `opinion cluster'. For larger values of $\varepsilon$, in accordance with Corollary~\ref{theorem:one_cautious_s_connected}, a consensus emerges, and this consensus opinion is the opinion of the opinion leader $C_1$ (viz., $m(\theta_1)=0.80$). As Fig.~\ref{fig:C_simu_bifur} indicates, this consensus begins to emerge for $\varepsilon>0.46$ (approx.). The dynamic nature of opinion exchange in the first three iterations is illustrated in Figs~\ref{fig:C_iteration_0}, \ref{fig:C_iteration_1}, and \ref{fig:C_iteration_2}. 

\textbf{\emph{Two Opinion Leaders.}} 
Here we replaced the two agents $\{R_1, R_7\}$ in Fig.~\ref{fig:A_simu_nx} by the cautiously updating agents $\{C_1, C_7\}$, respectively. Figs~\ref{fig:CC_nx_ori} and \ref{fig:CC_simu_bifur} show the corresponding network topology and bifurcation diagram for $\pmb{\pi}(\theta_1)$, respectively. In accordance with Corollary~\ref{theorem:two_cautious_s_connected}, no consensus is reached because the two opinion leaders $\{C_1, C_7\}$ possess different opinions. 

With the opinions of two opinion leaders being different, the number of opinion clusters created depends on the bound of confidence $\varepsilon$. For $0.31<\varepsilon<0.42$ (approx.), we observe two opinion clusters, the minimum number of clusters possible. We have achieved this by picking the agent BoEs carefully so that the network separates into two components, each with its own opinion leader, for the aforementioned values of $\varepsilon$. For these values, the network gets separated into two components $\{C_1, R_2, R_3, R_, R_5\}$ and $\{C_7, R_6\}$ because $\Vert\mc{E}_3-\mc{E}_6\Vert>\varepsilon$ and $\Vert\mc{E}_5-\mc{E}_7\Vert>\varepsilon$, for $0.31<\varepsilon<0.42$ (approx.). The ensuing network generates two opinion clusters at the opinions of the two opinion leaders $C_1$ and $C_7$. For larger values of $\varepsilon$, some (or all) receptively updating agents get influenced by both opinion leaders which creates different opinion clusters that are influenced by both opinion leaders. 

As asserted in Corollary~\ref{theorem:two_cautious_s_connected}, when the two opinion leaders possess different opinions, the receptively updating agents will reach a consensus if the matrices $C_k^{(1)}$ and $C_k^{(2)}$ in Definition~\ref{def:2-ODC} satisfy $\lambda_k^{(1)}C_k^{(1)}\mb{1}_{N_{C1}}=\lambda_k^{(2)}C_k^{(2)}\mb{1}_{N_{C2}},\,\forall k\in\mathbb{N}_{\geq 0}$. Fig.~\ref{fig:CC_nx_alt} shows a network topology which satisfies this condition, and the corresponding bifurcation diagram in Fig.~\ref{fig:CC_simu_bifur_alt} shows the emergence of a third opinion cluster around $m(\theta)_1=0.50$ (approx.) for $\varepsilon>0.43$ (approx.).

\subsubsection{Simulations with 100 Agents}
\label{sec:SimulationsWith100Nodes}

\begin{figure*}[htp]
  \subfigure[No opinion leaders.\label{fig:A_100}]{%
    \includegraphics[width=0.33\textwidth]{%
      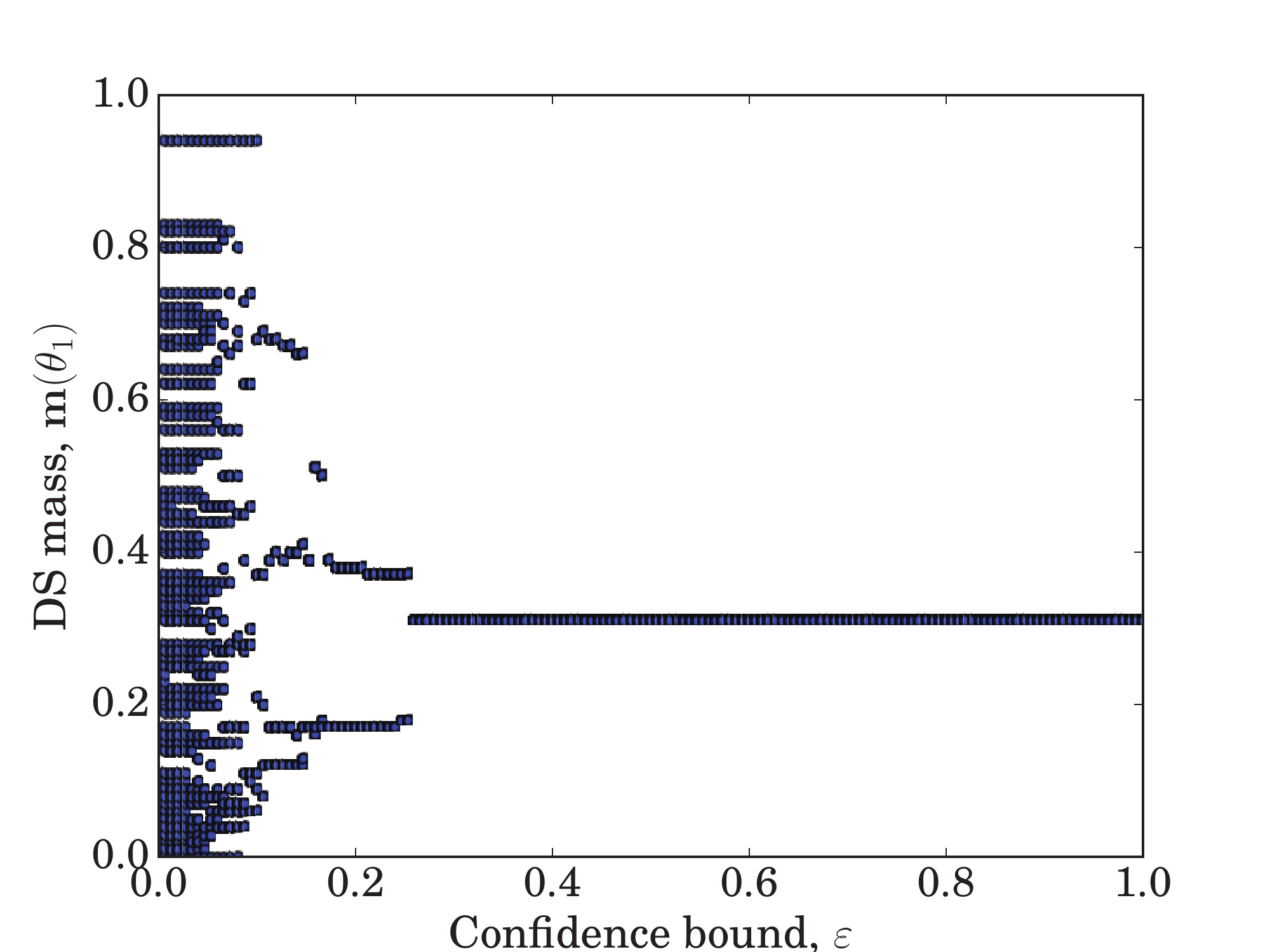}}
  \subfigure[One opinion leader.\label{fig:C_100}]{%
	\includegraphics[width=0.33\textwidth]{%
	  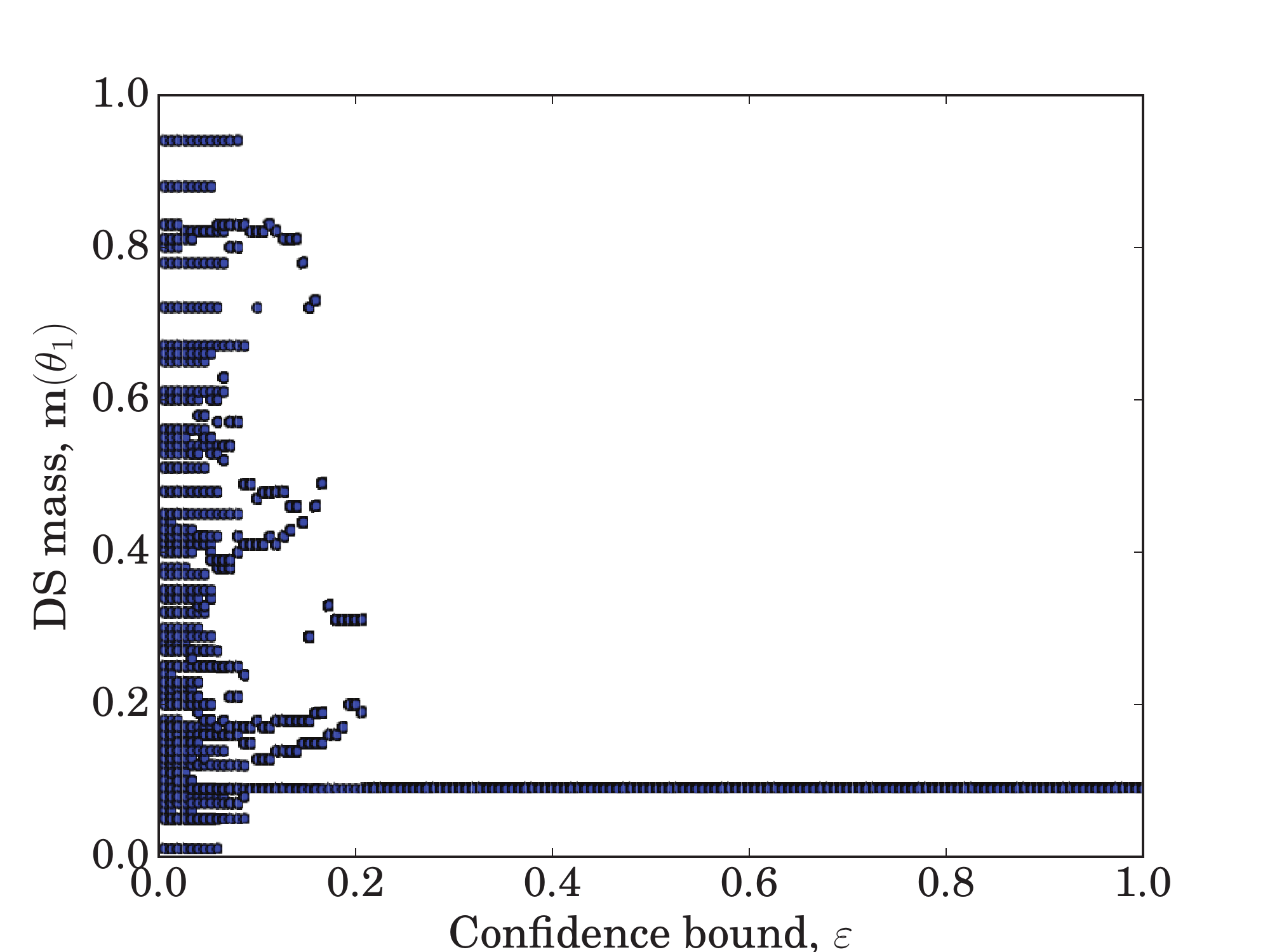}}
  \subfigure[Two opinion leaders.\label{fig:CC_100}]{%
    \includegraphics[width=0.33\textwidth]{%
      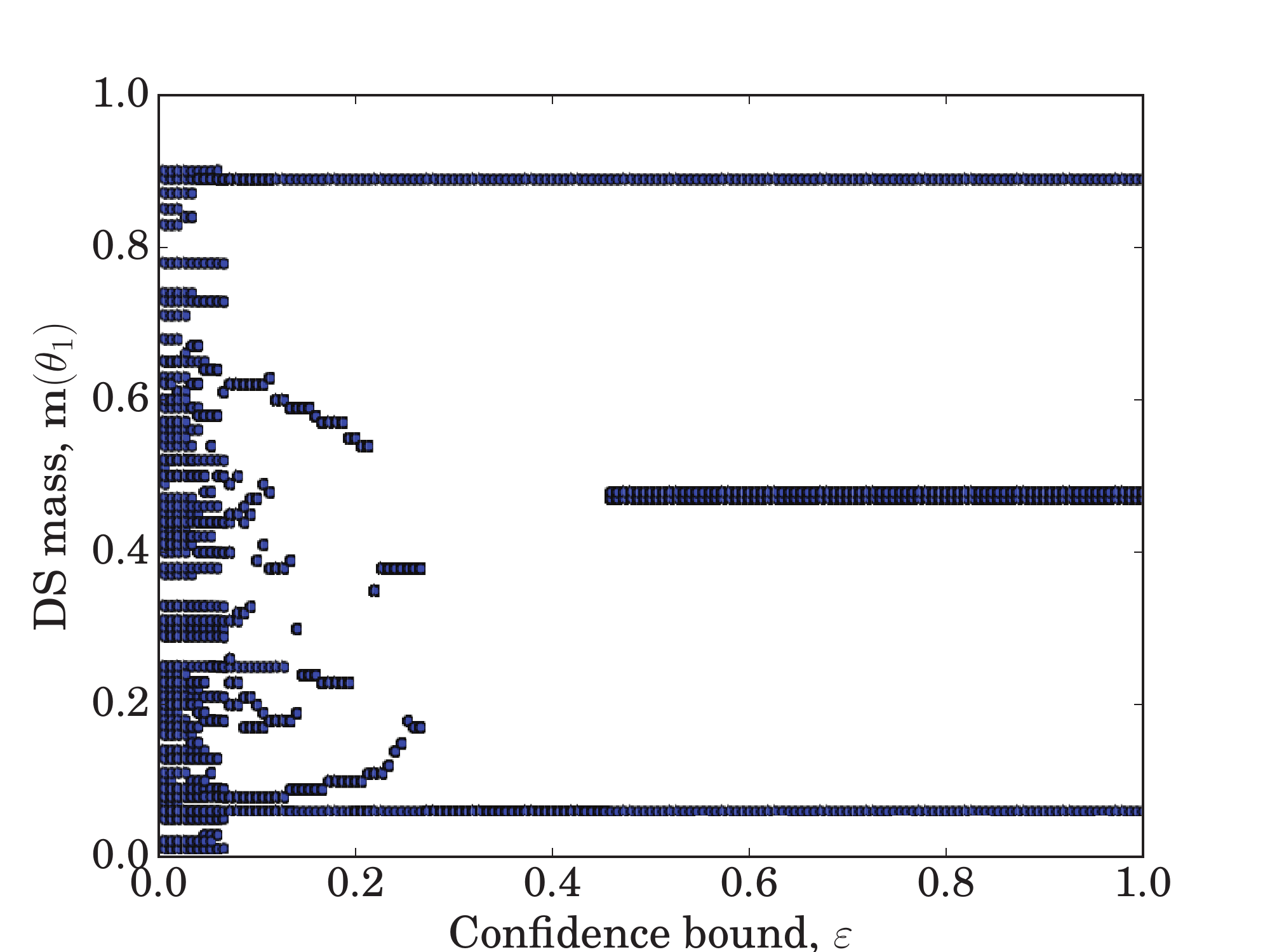}} \\
  \caption{Probabilistic agents: Simulation results for $100$ agents embedded in an Erd\H{o}s-R\'{e}nyi random graph with $p=0.10$ and agent BoEs sampled from $\text{Dir}(1,1,1)$. Consensus can be seen in Fig.~\ref{fig:A_100} and \ref{fig:C_100}, for $\varepsilon>0.26$ and $\varepsilon>0.21$, respectively.}
  \label{fig:100agents}
\end{figure*}  

\begin{figure*}[htp]
  \subfigure[No opinion leaders.\label{fig:DAOs_A7}]{%
    \includegraphics[width=0.33\textwidth]{%
      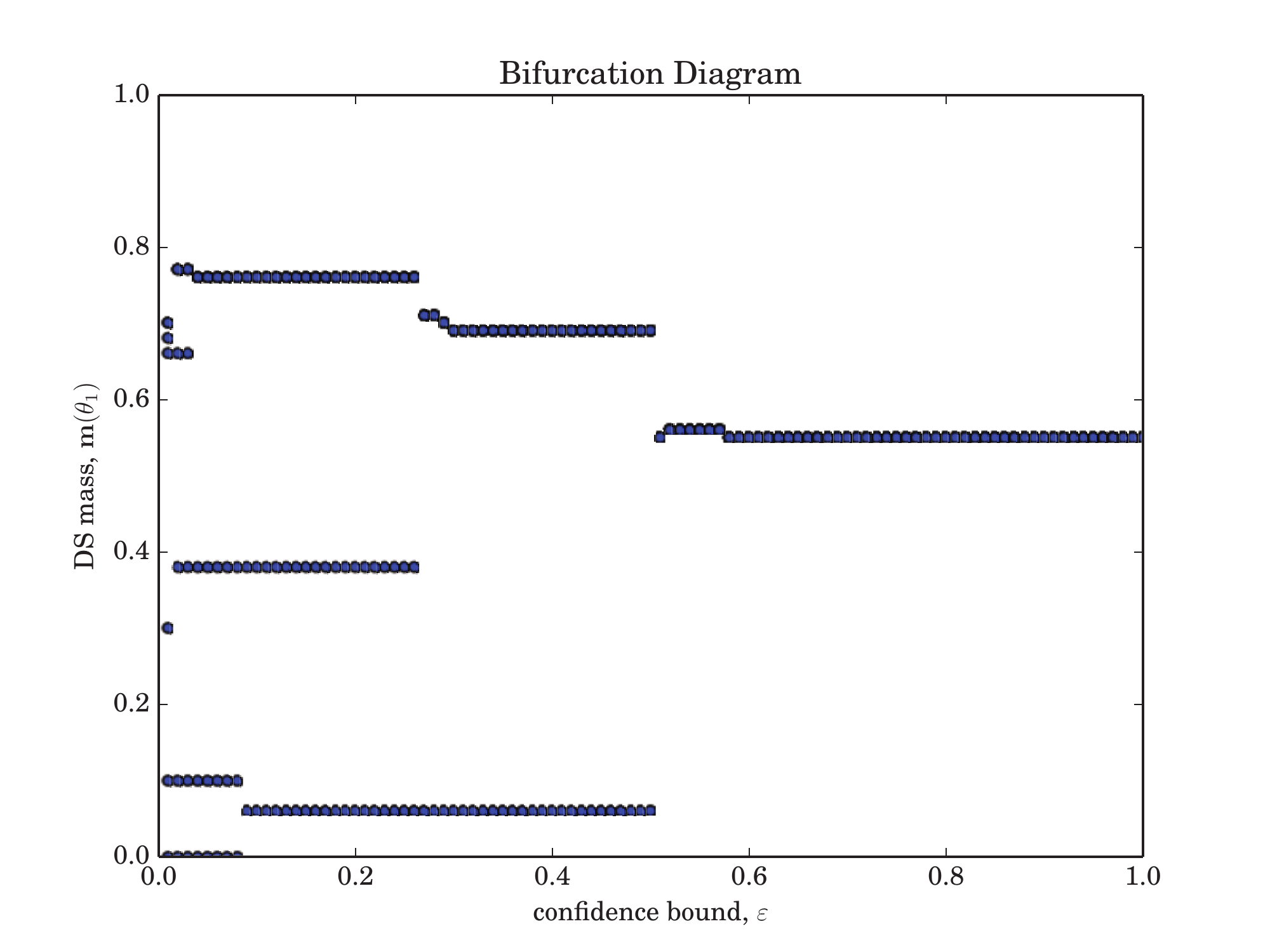}}
  \subfigure[One opinion leader.\label{fig:DAOs_C7}]{%
	\includegraphics[width=0.33\textwidth]{%
	  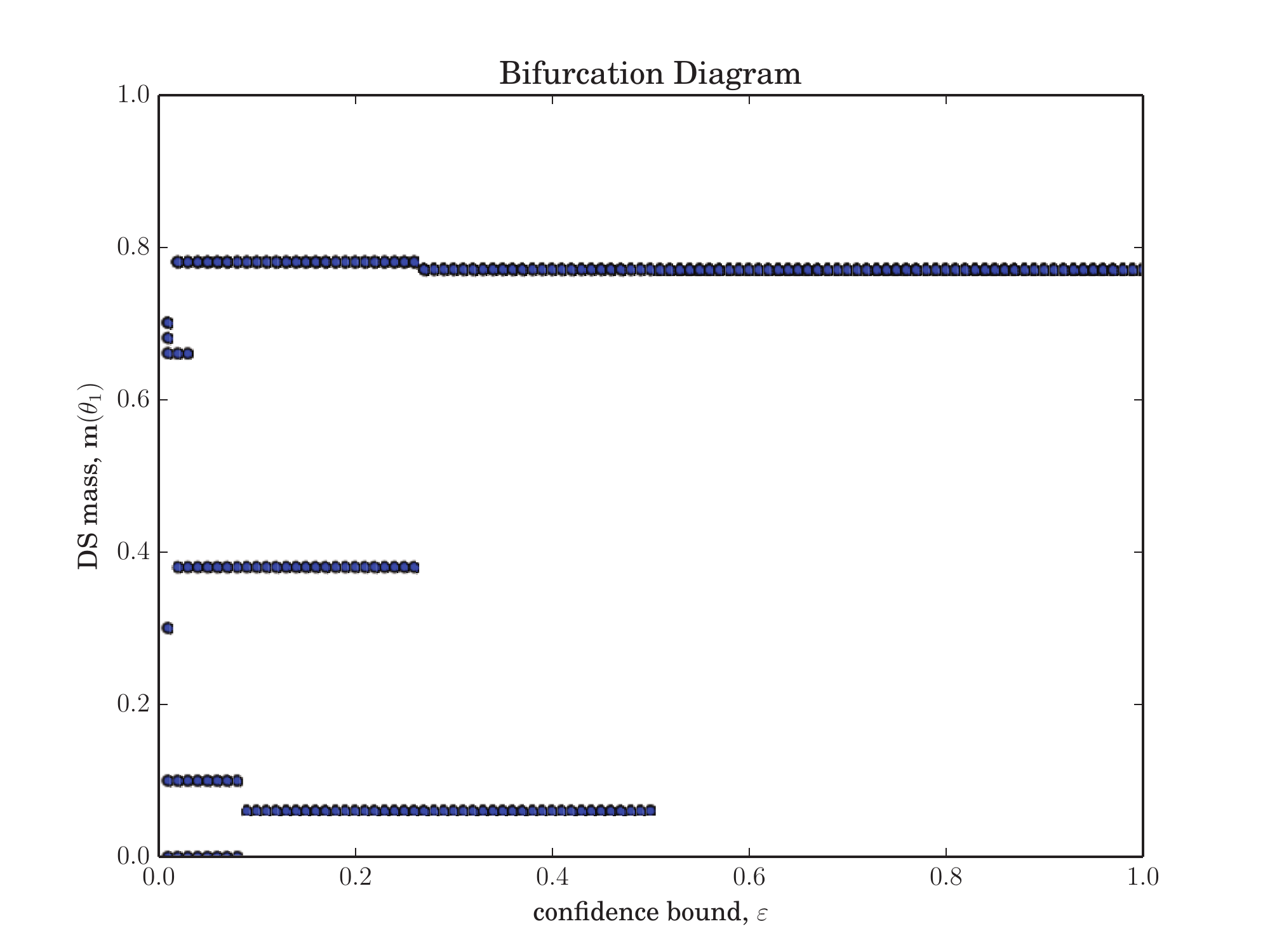}}
  \subfigure[Two opinion leaders.\label{fig:DAOs_CC7}]{%
    \includegraphics[width=0.33\textwidth]{%
      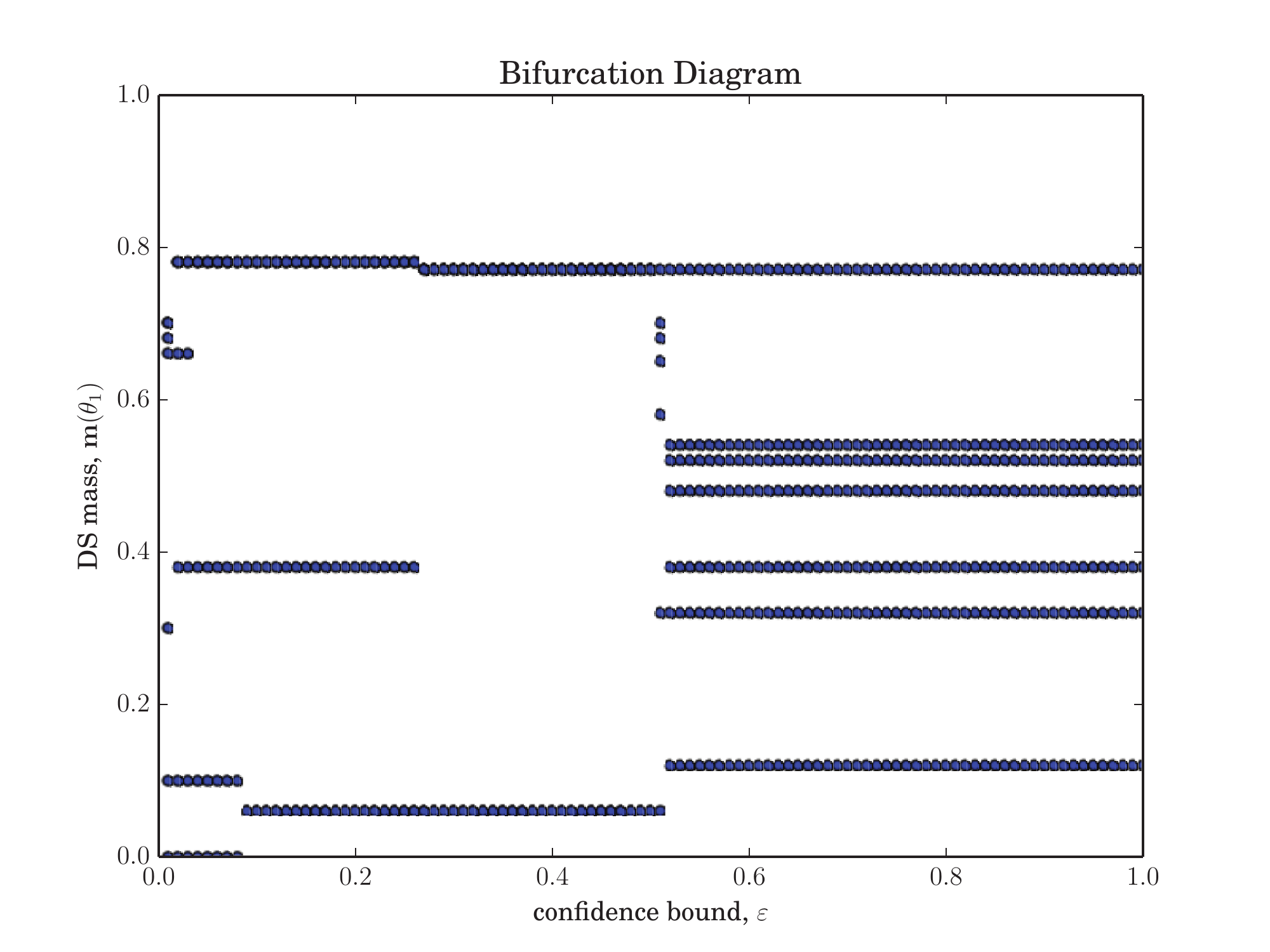}} \\
  \caption{Dirichlet agents: Simulation results for $7$ agents with no opinion leaders, one opinion leader and two opinion leaders embedded in the graphs in Figs~\ref{fig:A_simu_nx}, \ref{fig:C_simu_nx}, and \ref{fig:CC_nx_ori} respectively. Consensus can be seen in Fig.~\ref{fig:DAOs_A7} and \ref{fig:DAOs_C7}, for $\varepsilon>0.51$ and $\varepsilon>0.5$, respectively.}
  \label{fig:DAOs_7_agents}
\end{figure*} 
 
Here we embed $100$ agents in random graphs of 100 nodes generated using the Erd\H{o}s-R\'{e}nyi (ER) random graph model. As is well known, for an ER random graph, the phase transition for network connectivity occurs when the edge formation probability $p$ exceeds $\ln\,n/n$ \cite{Newman2012N}. With $n=100$, $\ln\,n/n=0.046$ and we used $p=0.10$ for generating all our random graphs. Moreover, every random graph was tested for connectedness at initialization. For sufficiently large values of $\varepsilon$, the graph $\mc{G}_k^{\dag}(\pmb{\varepsilon)}$ is therefore essentially the same as $\mc{G}_k$, and thus it is connected as well.

The BoE of each agent was sampled from the symmetric Dirichlet distribution $\text{Dir}(1,1,1)$, which is equivalent to a uniform distribution over the open standard 2-simplex \cite{Bishop2006PRML}. As Figs~\ref{fig:A_100} and \ref{fig:C_100} show, for $\varepsilon>0.26$ (approx.) and $\varepsilon>0.21$ (approx.), a consensus appears when there are no opinions leaders and when only one opinion leader is present, respectively. In accordance with Corollary~\ref{theorem:two_cautious_s_connected}, Fig.~\ref{fig:CC_100} shows that there is no consensus among the $100$ agents when the two opinion leaders have different opinions. 


\subsection{Dirichlet Agent Opinions}
\label{DAOs}


Here, we repeat the experiments conducted with the $7$-agent topologies in Section~\ref{sec:SimulationsWithSevenNodes} but with Dirichlet agent opinions. For all the agents, we kept the same mass vectors $\pmb{\pi}(\theta_2)_0$ and $\pmb{\pi}(\theta_3)_0$ as those in Section \ref{sec:SimulationsWithSevenNodes} while we used $\pmb{\pi}(\Theta)_0=0.1$; the remaining masses were assigned to $\pmb{\pi}(\theta_1)_0$. 

Fig.~\ref{fig:DAOs_7_agents} shows the corresponding bifurcation diagrams. As is evident, and in consistent with Corollary~\ref{theorem:one_cautious_s_connected_Dirichlet}, a consensus can be seen in Figs~\ref{fig:DAOs_A7} and \ref{fig:DAOs_C7} for $\varepsilon>0.51$ and $\varepsilon>0.5$, respectively. However, in Fig.~\ref{fig:DAOs_CC7}, there is no consensus among the agents. This is consistent with Corollary~\ref{theorem:two_cautious_s_connected_Dirichlet}(i) because the cautious agents do not possess the same converged opinion. The minimum number of opinion clusters appear for $0.26<\varepsilon<0.52$ (approx.). 

\begin{figure}[htpb]
  \centering    
   \includegraphics[width=0.30\textwidth]{%
     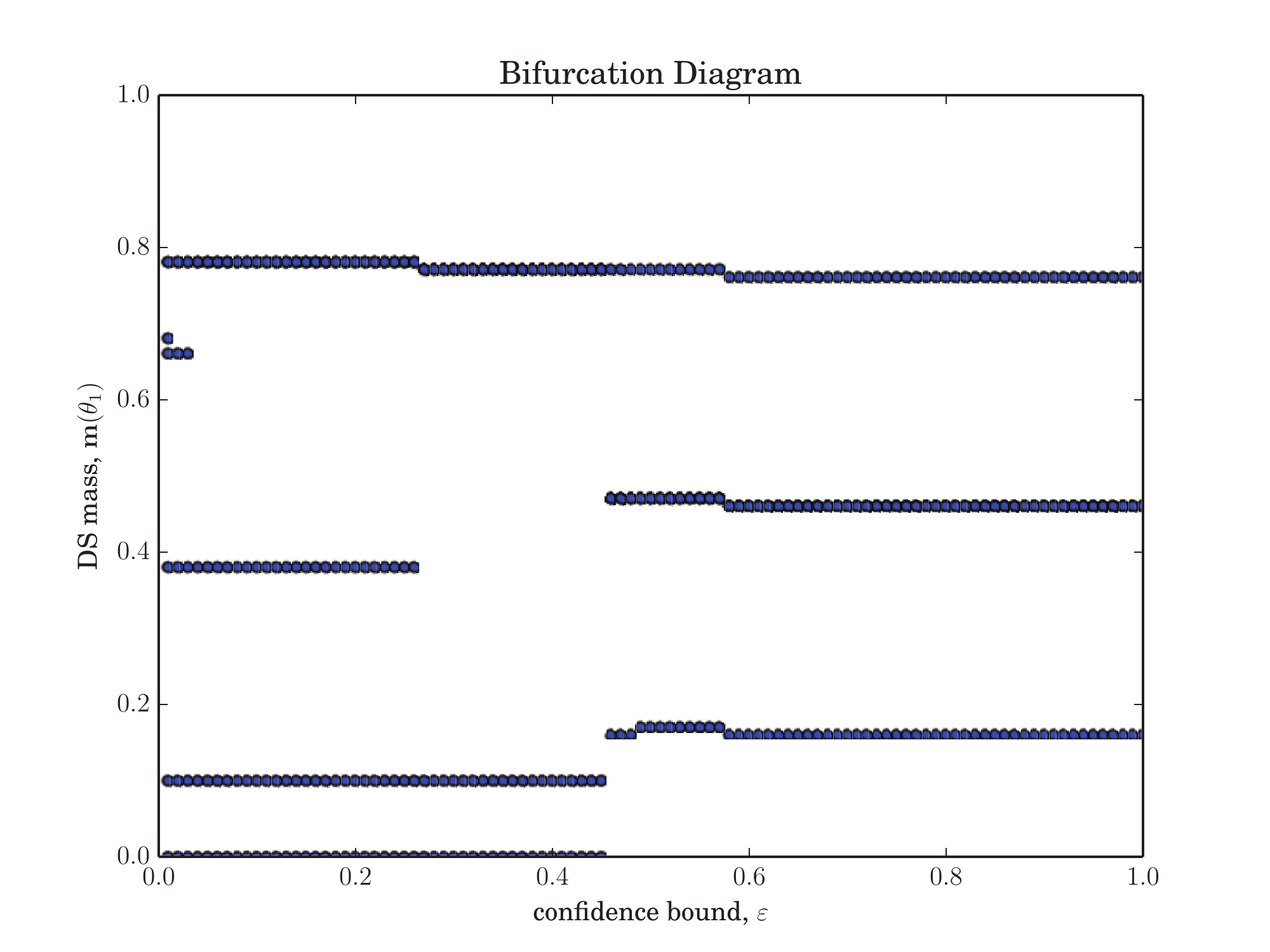}
   \caption{Dirichlet agents: Simulation results for  two opinion leaders ($C_1$ and $C_7$) and five receptively updating agents ($R_i,\,i\in\ol{2,6}$) embedded in the graph in Fig. \ref{fig:CC_nx_alt} generates a consensus among the receptively updating agents. This consensus appears for $\varepsilon> 0.46$ (approx.).} 
  \label{fig:DAOs_CC7_altered}
\end{figure}

Interestingly, even for higher values of $\varepsilon$, no consensus emerges even among the receptive agents. Indeed, one would expect that the receptive agents who are now less restrained to exchange opinions with their neighbors would form an opinion clusters of their own. In contrast, when the agents are embedded in the graph shown in Fig.~\ref{fig:CC_nx_alt}, a consensus emerges among the receptive agents for $\varepsilon> 0.46$ (approx.). See Fig.~\ref{fig:DAOs_CC7_altered}. This is because the graph topology in Fig.~\ref{fig:CC_nx_alt} satisfies the condition in Corollary~\ref{theorem:two_cautious_s_connected_Dirichlet}(ii). 


\begin{figure*}[htp]
  \subfigure[No opinion leaders.\label{fig:ds_A_7}]{%
    \includegraphics[width=0.33\textwidth]{%
      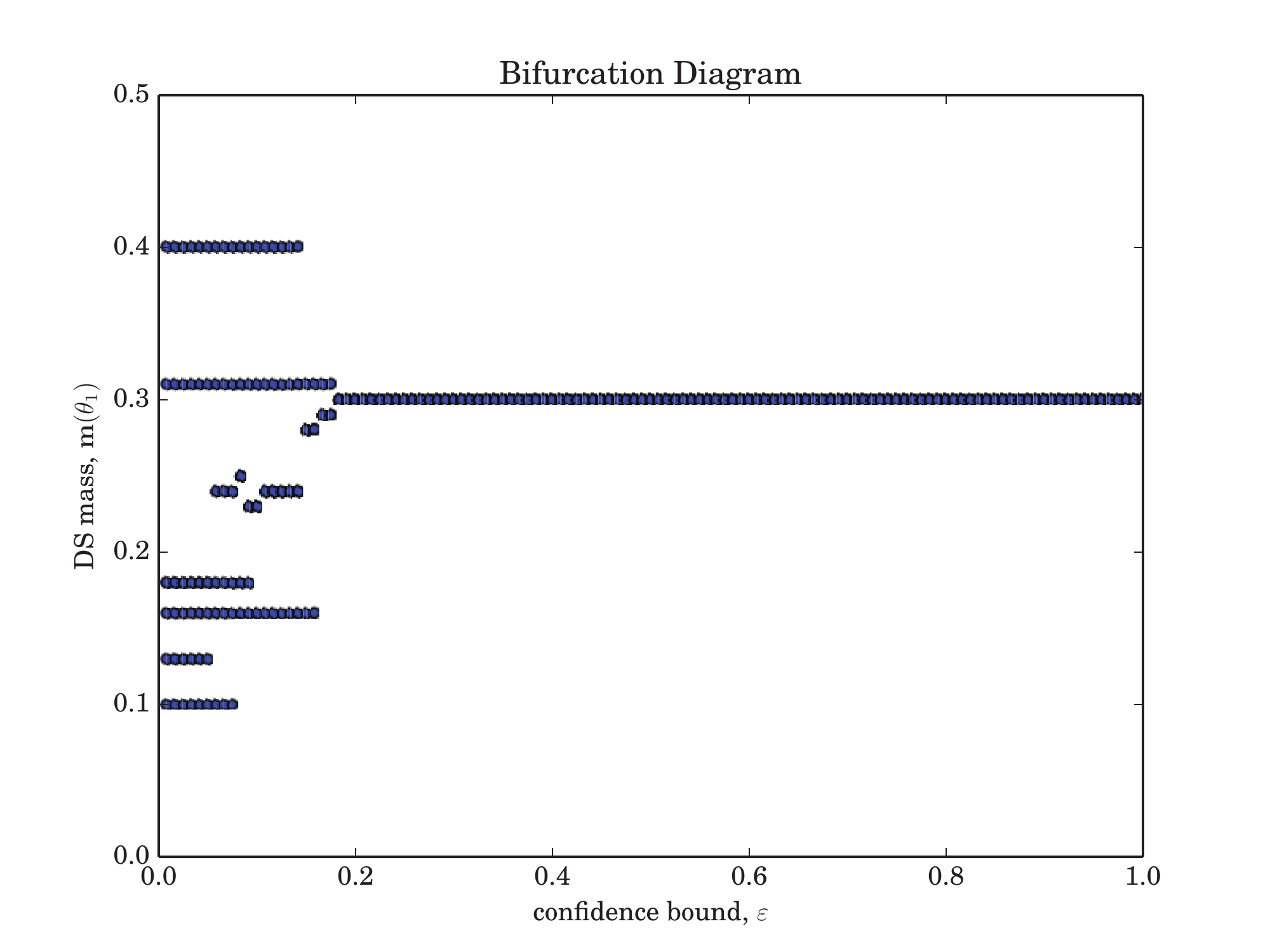}}
  \subfigure[One opinion leader.\label{fig:ds_C_7}]{%
	\includegraphics[width=0.33\textwidth]{%
	  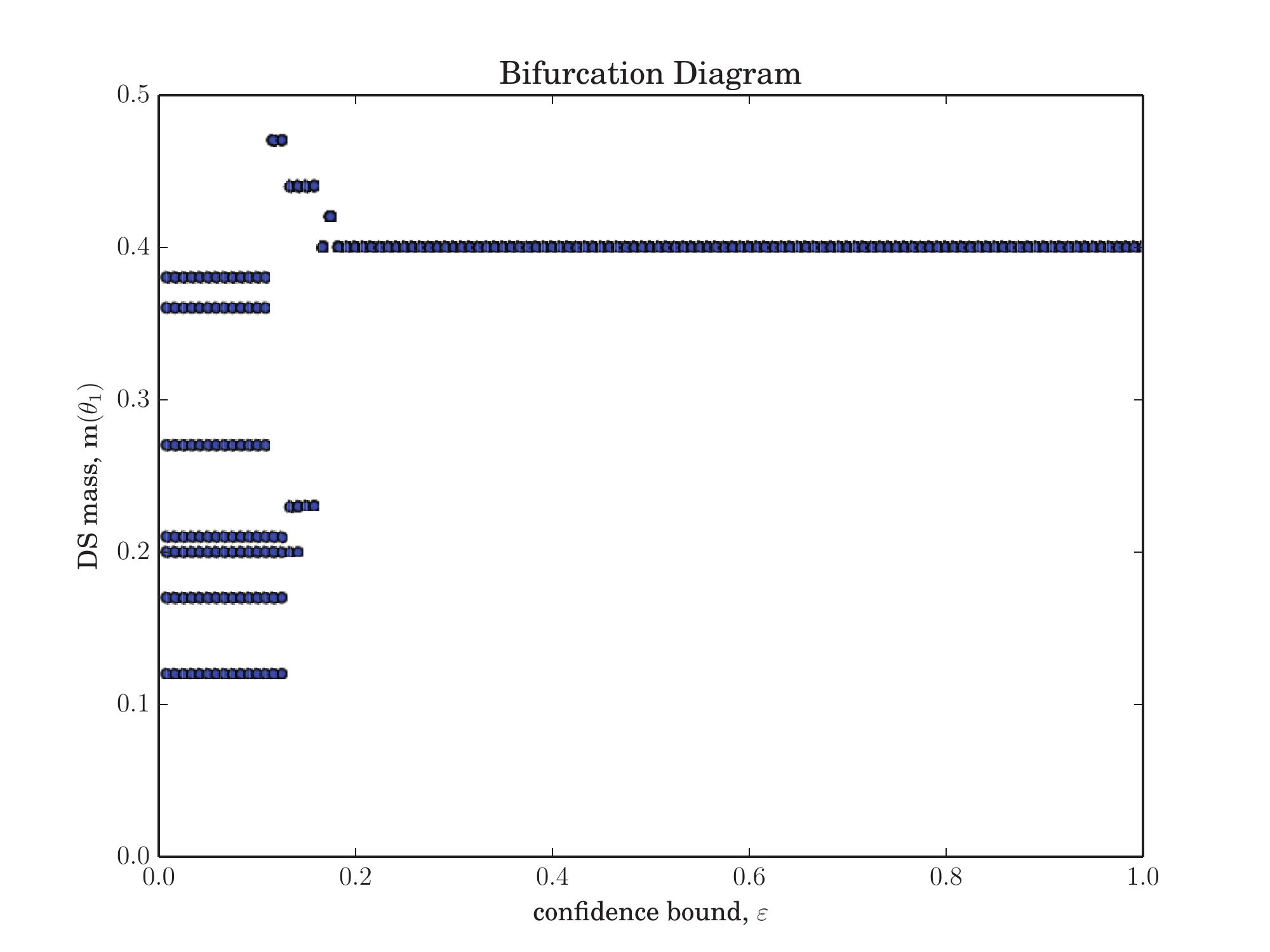}}
  \subfigure[Two opinion leaders.\label{fig:ds_CC_7}]{%
    \includegraphics[width=0.33\textwidth]{%
      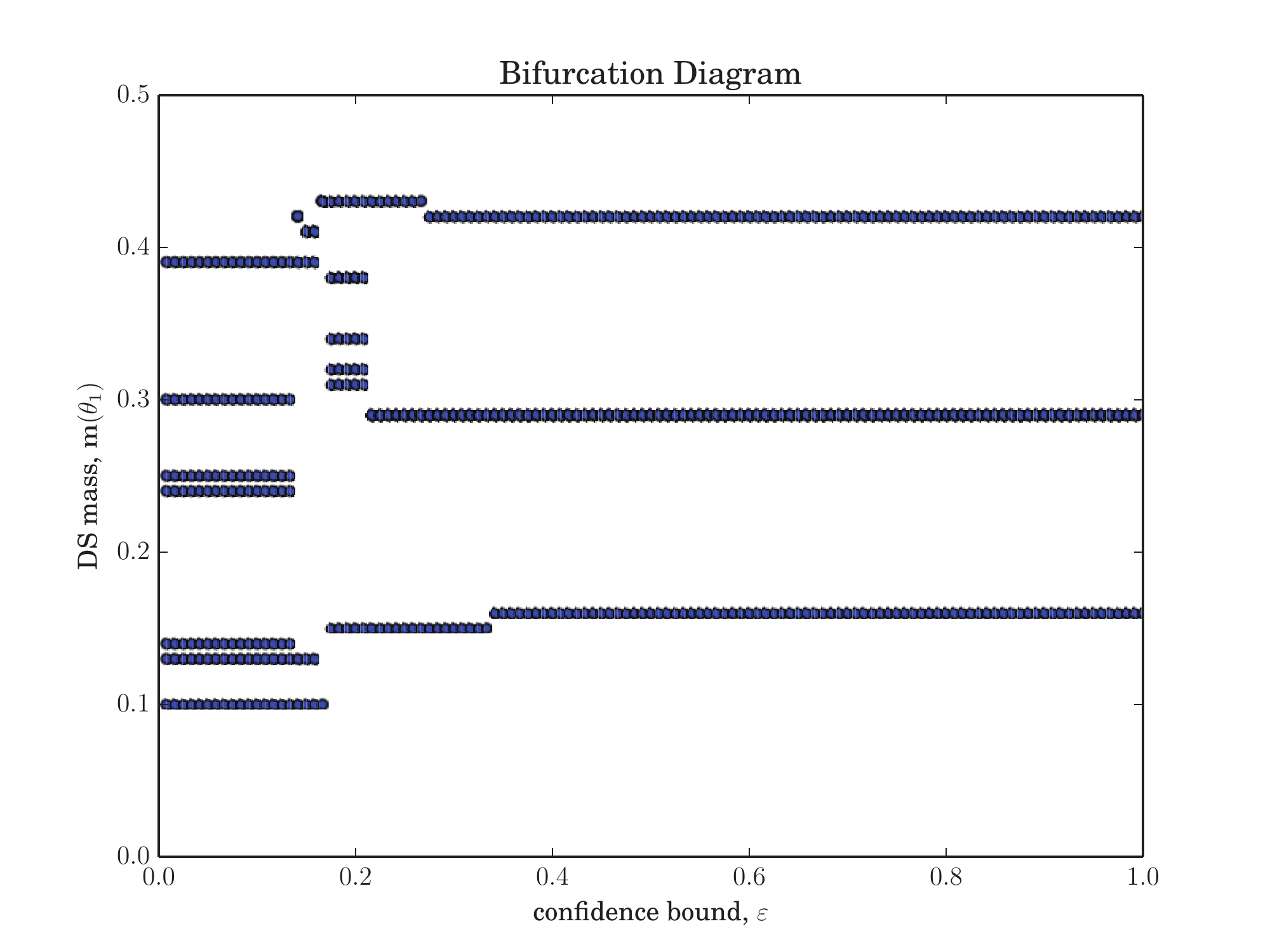}} \\
  \caption{General DST agents: Simulation results for seven receptively updating agents  embedded within the network topology in \ref{fig:CC_nx_alt} and DST mass values sampled from $\text{Dir}(4,4,4,2,2,2,1)$. A consensus appears for the no opinion leader and one opinion leaders cases in Fig.~\ref{fig:ds_A_7} and \ref{fig:ds_CC_7} for $\varepsilon>0.18$ (approx.)  and $\varepsilon>0.17$ (approx.), respectively.}
  \label{fig:ds7agents}
\end{figure*}  

\subsection{General DST Agent Opinions}
\label{DSTAOs}


\subsubsection{Simulations with Seven Agents}
\label{sec:SimulationsWithSevenNodes2}

In this study, we assigned random DST mass assignments for the seven agents embedded within the topology in Fig.~\ref{fig:CC_nx_alt}. For this purpose, we utilized the Dirichlet distribution which has been widely employed in opinion modeling \cite{Chaloner1987, Regazzini1999, Vazquez2014}. In particular, for each agent in each trial, the DST masses for $\theta_1$, $\theta_2$, $\theta_3$, $(\theta_1\theta_2)$, $(\theta_1\theta_3)$, $(\theta_2\theta_3)$, and $\Theta=(\theta_1\theta_2\theta_3)$ were sampled from the Dirichlet distribution $\text{Dir}(4,4,4,2,2,2,1)$. Figs~\ref{fig:ds_A_7}, \ref{fig:ds_C_7}, and \ref{fig:ds_CC_7} show the bifurcation diagrams when the network contains no opinion leaders, one opinion leader, and two opinion leaders, respectively. 

As Fig.~\ref{fig:ds_A_7} shows, with no opinion leaders, a consensus appears for $\varepsilon>0.18$ (approx.).  Fig.~\ref{fig:ds_C_7} shows bifurcation diagram when only one opinion leader is present, and we can see that a consensus appears for $\varepsilon>0.17$ (approx.). As Fig.~\ref{fig:ds_CC_7} shows, with two opinion leaders, no consensus is reached among the agents. However, it is interesting to note that three opinion clusters emerge for $\varepsilon>0.21$ (approx.). 

As mentioned earlier, the main reason for utilizing the DST framework for capturing agent opinions is its ability to capture the types of uncertainties and the nuances that are characteristic of agent states and opinions. DST agent opinions can also generate new emergent behavior which cannot be captured via probabilistic agents. For example, consider 7 receptive agents embedded within the topology in Fig.~\ref{fig:A_simu_nx}. The initial opinions and the converged opinions for $\varepsilon=0.30$ appear in Table~\ref{table:converged_opinions}. Notice that two opinion clusters have emerged: the first cluster formed by $\{R_1,\, \ldots,\, R_5\}$ converge to the probabilistic opinion $\{m_1^*(\theta_1),\, m_1^*(\theta_2),\, m_1^*(\theta_3)\}=\{0.63,\, 0.19,\, 0.18\}$; the second cluster formed by $R_6$ and $R_7$ converge to the general DST opinion $\{m_2^*(\theta_1),\, m_2^*(\theta_2,\, \theta_3)\}=\{0.15,\, 0.85\}$ which allows no further `refinement' between the singletons $\theta_2$ and $\theta_3$. Such emergent behavior is qualitatively different than what appears in prior models \cite{Deffuant2000, Krause2000, Hegselmann2002, Weisbuch2004, Lorenz2007}.

\begin{table}[htbp] 
  \centering
  \caption{Initial and Converged Opinions (with $\varepsilon=0.30$)}
  \begin{tabular}{c rrr rr} 
    \hline
    \textbf{Agent}
      & \multicolumn{5}{c}{\tb{DST Mass Values}} \\
    \cline{2-6}
    {}
      & $\theta_1$ & $\theta_2$ & $\theta_3$ 
      & $(\theta_1,\, \theta_2)$ & $(\theta_2,\, \theta_3)$ \\ 
    \hline\hline
    \multicolumn{6}{l}{\tb{Initial Opinions:}} \\
    $R_1$
      & 0.60 & 0.10 & 0.10 
      & 0.10 & 0.10 \\
    $R_2$
      & 0.62 & 0.11 & 0.04 
      & 0.11 & 0.12 \\
    $R_3$
      & 0.51 & 0.12 & 0.05 
      & 0.12 & 0.20	\\
    $R_4$
      & 0.57 & 0.15 & 0.03 
      & 0.15 & 0.10	\\
    $R_5$
      & 0.60 & 0.10 & 0.10 
      & 0.10 & 0.10	\\
    $R_6$
      & 0.10 & -- & -- 
      & -- & 0.90 \\
    $R_7$
      & 0.20 & -- & -- 
      & -- & 0.80 \\
    \hline
    \multicolumn{6}{l}{\tb{Converged Opinions (with $\varepsilon=0.30$):}} \\
    $\{R_1,\, \ldots,\, R_5\}$
      & 0.63 & 0.19	& 0.18
      &	-- & --- \\
    $\{R_6,\, R_7\}$
      & 0.15 & -- &	--
      &	-- & 0.85 \\
    \hline
  \end{tabular}
  \label{table:converged_opinions}
\end{table}


\section{Conclusion}


In this paper, we use the DST framework for representing agent opinions and explore the formation of consensus and opinion clusters when agents residing within a network exchange and update their opinions. In particular, we explore the effect that opinion leaders have on these processes. Our opinion model accounts for aspects from SJT and possesses the ability to capture a wider variety of uncertainties and nuances in agent opinions, an advantage inherited from its DST basis. Theoretical analysis, which focuses on probabilistic and Dirichlet agent opinions, provides conditions for the emergence of consensus and opinion clusters in the presence of opinion leaders. Our results show that a consensus can be formed when the number of opinion leaders is no more than one and with a sufficiently high bound of confidence of the agents. With two or more opinion leaders possessing different opinions, no consensus can be reached in general. We also explore the conditions for opinion cluster formation among the opinion followers. 

Our current work involves extending this theoretical analysis to scenarios where agent opinions are captured via more general DST BoEs, which may require recourse to tools from paracontractions theory \cite{Wickramarathne2014JoSTSP}.  It is also noteworthy that we have taken all agents to possess the identical bound of confidence value. When this is not the case, the opinion exchange mechanism itself would be directional (because an agent with a lower bound of confidence may update itself from its neighbor agent who may not update itself because of a higher bound of confidence value). An interesting future research problem is the study of networked agents whose bounds of confidence values are different. Another interesting issue to be addressed is the assessment of the convergence speed of our algorithms \cite{Olshevsky2009}.




\balance



\end{document}